\documentclass{article}  

\usepackage{amsmath,amssymb,amsfonts}

\title{\bf Conformal geodesics on vacuum space-times}
\author{Helmut Friedrich\\ 
Max-Planck-Institut f\"ur Gravitationsphysik\\
Am M\"uhlenberg 1\\
14476 Golm\\
Germany}

\begin{document}
\maketitle

\begin{abstract}
We discuss properties of conformal geodesics on general, vacuum, and
warped product space-times and derive a system of conformal deviation
equations. The results are used to show how to construct on the
Schwarzschild-Kruskal space-time global conformal Gauss coordinates which
extends smoothly and without degeneracy to future and past null
infinity. 
\end{abstract}

\newpage

\section{Introduction}

Geometrically defined coordinate systems are convenient in the
local analysis of fields but they often degenerate on larger
domains. It is well known, for instance, that Gauss coordinates are of
limited use because the underlying geodesics tend to develop caustics. In
\cite{friedrich:schmidt} conformal Gauss coordinates have been introduced
and successfully employed in local studies of conformally rescaled
space-times. Here the time-like coordinate lines are generated by
conformal geodesics, curves which are associated with
conformal structures in a similar way as geodesics are associated with
metrics. In this case the danger of a coordinate break-down is even more
severe, because the conformal geodesics generating the system can not
only develop envelopes or intersections, as occur in caustics
of metric geodesics, but they may even become tangent to each other.
However, it will be shown in this article that the structure responsible
for this difficulty also has useful aspects.

Subsequent analyses of conformal Gauss coordinates revealed that
conformal geode\-sics possess various remarkable properties. In
\cite{friedrich:AdS} they have been studied systematically in the context
of the conformal field equations. Though this at first introduces 
additional complications into the equations because it requires
the use of Weyl connections, it leads to unexpected simplifications. 
While the Bianchi equation satisfied by the rescaled conformal Weyl
tensor, which is given by $d^i\,_{jkl} = \Theta^{-1}\,C^i\,_{jkl}$ where
$\Theta$ denotes the conformal factor, always implies hyperbolic reduced
systems of partial differential equations, the generalized conformal field
equations (in vacuum, possibly with a cosmological constant) admit in the
new gauge hyperbolic reductions in which the frame coefficients, the
connection coefficients and the components of the conformal Ricci tensor
are governed by equations of the form
\[
\partial_{\tau}\,e^{\mu}\,_k = \ldots,\,\,\,\,\,\,\,\,\,
\partial_{\tau}\,\hat{\Gamma}_i\,^j\,_k = \ldots,\,\,\,\,\,\,\,\,\,
\partial_{\tau}\,\hat{R}_{jk} = \ldots,
\] 
where $\tau$ denotes the time variable in the given gauge and the right
hand sides are given by algebraic functions of the unknowns $e^{\mu}\,_k$,
$\hat{\Gamma}_i\,^j\,_k$, $\hat{R}_{jk}$, $d^i\,_{jkl}$ and known
functions on the solution manifold. 

Moreover, associated with a conformal geodesic is a function along that
curve, the {\it conformal factor}, which is determined up to a constant
factor that can be fixed on a given initial hypersurface. This function
necessarily has zeros at points where the conformal geodesic crosses
null infinity. It turns out that for given data on the initial
hypersurface this function can be determined explicitly and acquires a
form which allows us to prescribe to null infinity by a suitable choice of
initial data a finite coordinate location. 

In \cite{friedrich:AdS} these facts are used to provide a rather
complete discussion of Anti-de Sitter-type solutions, including a
characterization of these solutions in terms of initial and boundary data.
In \cite{friedrich:i0} these properties form a basic ingredient of a
detailed investigation of the behaviour of asymptotically flat solutions
near space-like and null infinity under certain assumptions on the Cauchy
data for the solutions. In particular, the {\it cylinder at space-like
infinity} introduced in \cite{friedrich:i0} to remove the conformal
singularity at space-like infinity is obtained as a limit set of conformal
geodesics. The results obtained there strongly suggest that even under
the most stringent smoothness assumptions on the conformal data near
space-like infinity the solutions will in general develop logarithmic
singularities at null infinity. However, they also suggest that these
singularities can be avoided if the data satisfy certain {\it regularity
conditions} at space-like infinity. 

The conversion of these conjectures into facts requires the proof of a
certain existence result for the {\it regular finite initial value problem
at space-like infinity} formulated in \cite{friedrich:i0}. While the basic
difficulty of this proof has nothing to do with the specific features of
the underlying conformal Gauss system, only the proof will show that the
latter is doing what we expect it to do: to extend near space-like
infinity smoothly and without degeneracy to null infinity if null
infinity admits a smooth differentiable structure at all.  

An independent proof that conformal Gauss systems behave this way on
general space-times is difficult, because it requires the information on
the asymptotic structure of the solution which we first hope to
obtain by completing the analysis of \cite{friedrich:i0}. The
question whether conformal Gauss systems can be used globally is even
more difficult. 

In the context of the hyperboloidal initial value problem with smooth
initial data it can be shown that these coordinates remain good for a
while and that they can in fact be globally defined for data sufficiently
close to Minkowskian hyperboloidal data. However, nothing is known for
data which deviate strongly from Minkowskian once.  There remains the
possibility to study the behaviour of conformal Gauss systems on specific
solutions. It has been shown in \cite{friedrich:i0} that there exist
conformal Gauss systems on the Schwarzschild space-time  which smoothly
cover  neighbourhoods of space-like infinity including parts of null
infinity. 

One may be tempted to think that this is only possible due to the weakness
of the field in that domain (although it doesn't look weak in the
conformal picture) while in regimes of strong curvature a break-down of
the coordinates will occur. It turns out that this is not true. 
It is the main results of this article that there exist conformal Gauss
coordinates on the Schwarzschild-Kruskal space-time which smoothly cover
the whole space-time and which extend smoothly (in fact analytically) and
without degeneracy through null infinity.

This raises hopes that conformal Gauss systems stay regular under much
wider circumstances than expected so far and that we may be able to
exploit the simplicity of the reduced equations and the fact that 
conformal Gauss systems provide by themselves the conformal
compactification in time directions under quite general assumptions. 
How robust these systems really are, whether they will
stay regular under non-linear perturbations of the Schwarschild-Kruskal
space, remains to be seen. Though a general analytical investigation of
these equations appears difficult at present, important information
about them may be obtained by numerical calculations. Conversely, 
there are good reasons to believe that it should be possible to give
strong analytical support to numerical work based on the use of conformal
Gauss systems because the latter are amenable to a geometric analysis. 

We begin the article by discussing properties of conformal geodesics on
general space-times and then specialize to vacuum space-times, using also
some of the results which have been obtained already in the articles
referred to above. For later application we work out the conformal
geodesic equations on warped product space-times. Various features of
conformal Gauss systems are then illustrated by explicit examples on
Minkowski space-time. 

After specializing further to warped product vacuum space-times a
conformal Gauss system on the Schwarzschild-Kruskal space-time will be
analysed.  Expressions for the conformal geodesics determined by suitably
chosen initial data are derived in terms of elliptic and theta functions.
It is shown that the curves extend smoothly through null infinity and that
the associated conformal factor defines a smooth conformal extension if
the congruence is regular. We do not attempt to derive the explicit
expression of the rescaled metric in terms of the new coordinates, because
this cannot be done anyway in more general situations. Instead, the
regularity of the coordinate system and the conformal extension is
established by analysing a system of equations which is the analogue
for conformal geodesics of the Jacobi equation for metric geodesics. While
specific features of the Schwarzschild solution are used here, it is
clear that similar techniques apply in more general
cases. We end the article by indicating as a possible application 
the numerical calculation of entire asymptotically flat
vacuum solutions.

\section{Conformal geodesics on pseudo-Riemannian manifolds}

Before we introduce conformal geodesics we  recall a few concepts
and formulae of conformal geometry.  On a pseudo-Riemannian manifold
$(M,
\tilde{g})$ of dimension
$n
\ge 3$ we consider two operations preserving the conformal
structure defined by $\tilde{g}$: (i) conformal rescalings of the metric
\begin{equation}
\label{conresc}
\tilde{g}_{\mu \nu} \rightarrow g_{\mu \nu} = \Omega^2\,\tilde{g}_{\mu \nu},
\end{equation}
with smooth conformal factor $\Omega > 0$, (ii) transitions 
$\nabla \rightarrow \hat{\nabla}$ of the Levi-Civita connection 
$\nabla$ of a metric $g$ in the conformal class of $\tilde{g}$ into
(torsion free) {\it Weyl connections} $\hat{\nabla}$ with respect
to $\tilde{g}$. In terms of the Christoffel symbols $\Gamma$ and
the connection coefficients $\hat{\Gamma}$, defined by 
$\nabla_{\partial_{\mu}}\,\partial_{\nu} = 
\Gamma_{\mu}\,^{\rho}\,_{\nu}\,\partial_{\rho}$ and 
$\hat{\nabla}_{\partial_{\mu}}\,\partial_{\nu} = 
\hat{\Gamma}_{\mu}\,^{\rho}\,_{\nu}\,\partial_{\rho}$ respectively,
the transition is described by
\begin{equation}
\label{WGtraf}
\Gamma_{\mu}\,^{\rho}\,_{\nu} \rightarrow
\hat{\Gamma}_{\mu}\,^{\rho}\,_{\nu} = 
\Gamma_{\mu}\,^{\rho}\,_{\nu} + S(f)_{\mu}\,^{\rho}\,_{\nu},
\end{equation}
\[ 
S(f)_{\mu}\,^{\rho}\,_{\nu} \equiv
\delta^{\rho}\,_{\mu}\,f_{\nu} 
+ \delta^{\rho}\,_{\nu}\,f_{\mu} 
+ g_{\mu \nu}\,g^{\rho \lambda}\,f_{\lambda},
\]
with a smooth 1-form $f$. For given $f$ the difference tensor 
$S(f)$ depends only on the conformal structure of $g$. The
transformation of the Levi-Civita connection $\tilde{\Gamma}$ of
$\tilde{g}$ under
\ref{conresc} is a special case of (\ref{WGtraf}) in which the 1-form
is exact  
\begin{equation}
\label{mGtraf}
\tilde{\Gamma}_{\mu}\,^{\rho}\,_{\nu} 
\rightarrow \Gamma_{\mu}\,^{\rho}\,_{\nu} = 
\tilde{\Gamma}_{\mu}\,^{\rho}\,_{\nu} + S(f)_{\mu}\,^{\rho}\,_{\nu}
\quad\mbox{with}\quad f = \Omega^{-1}\,d\,\Omega.
\end{equation}
Conversely, if the 1-form $f$ is closed (\ref{WGtraf}) arises locally
from a rescaling of $g$ with a suitable conformal factor. By
(\ref{WGtraf}) we have   
\begin{equation}
\label{wdg}
\hat{\nabla}_{\rho}\,g_{\mu \nu} = - 2\,f_{\rho}\,g_{\mu \nu},
\end{equation}
It follows from this that $\hat{\nabla}$ preserves
the conformal structure of $\tilde{g}$ in the sense that with a
function $\theta > 0$ satisfying on a given curve $x(\tau)$ in
$M$
\begin{equation}
\label{wpcofac}
\hat{\nabla}_{\dot{x}}\,\theta = \theta <f, \dot{x}>,
\end{equation}
the metric $\theta^2 g_{\mu \nu}$ is parallely
transported along $x(\tau)$ with respect to $\hat{\nabla}$. 

The curvature tensors of the connections $\hat{\nabla}$ and
$\nabla$, defined by 
$(\hat{\nabla}_{\lambda}\,\hat{\nabla}_{\rho} - 
\hat{\nabla}_{\rho}\,\hat{\nabla}_{\lambda})\,X^{\mu} =
\hat{R}^{\mu}\,_{\nu \lambda \rho}\,X^{\nu}$ 
etc., are related by 
\begin{equation}
\label{curvWGtraf}
\hat{R}^{\mu}\,_{\nu \lambda \rho} - R^{\mu}\,_{\nu \lambda \rho} =
2\,\{\nabla_{[\lambda}\,S_{\rho]}\,^{\mu}\,_{\nu}
+ S_{\delta}\,^{\mu}\,_{[\lambda}\,S_{\rho]}\,^{\delta}\,_{\nu}\}
\end{equation}
were indices are raised or lowered with respect to $g$.
The tensor
\[
\hat{L}_{\mu \nu} = \frac{1}{n - 2}\,\{\,\hat{R}_{(\mu \nu)}
- \frac{n - 2}{n}\,\hat{R}_{[\mu \nu]}
- \frac{1}{2\,(n - 1)}\,g_{\mu \nu}
\,\hat{R}\,\},
\]
which occurs in the decomposition 
$\hat{R}^{\mu}\,_{\nu \lambda \rho} =
2\,\{g^{\mu}\,_{[\lambda}\,\hat{L}_{\rho]\nu}  -
g^{\mu}\,_{\nu}\,\hat{L}_{[\lambda \rho]}  -
\,g_{\nu[\lambda}\,\hat{L}_{\rho]}\,^{\mu}\} 
+ C^{\mu}\,_{\nu \lambda \rho}$ 
of the curvature tensor of $\hat{\nabla}$ into its trace parts and
the trace-free, conformally invariant conformal Weyl tensor 
$C^{\mu}\,_{\nu \lambda \rho}$, is related 
to the tensor
\[
L_{\mu \nu} = \frac{1}{n - 2}\,\{\,R_{\mu \nu}
- \frac{1}{2\,(n - 1)}\,R\,g_{\mu \nu}\,\}, 
\]
by the equation
\begin{equation}
\label{WLtens}
\nabla_{\nu}\,f_{\mu} - f_{\mu}\,f_{\nu}
+ g_{\mu \nu}\,\frac{1}{2}\,f_{\lambda}\,f^{\lambda}
= L_{\mu \nu} - \hat{L}_{\mu \nu}.
\end{equation}

\subsection{Conformal geodesics}

A conformal geodesic associated with the conformal structure
defined  by $\tilde{g}$ is given by a curve $x(\tau)$ in $M$ and a
1-form
$b(\tau)$ along $x(\tau)$ such that the equations
\begin{equation}
\label{acgxequ}
(\tilde{\nabla}_{\dot{x}}\dot{x})^{\mu}
+ S(b)_{\lambda}\,^{\mu}\,_{\rho}\,\dot{x}^{\lambda}\,\dot{x}^{\rho} = 0, 
\end{equation}
\begin{equation}
\label{bcgbequ}
(\tilde{\nabla}_{\dot{x}}b)_{\nu} - \frac{1}{2}\, 
b_{\mu}\,S(b)_{\lambda}\,^{\mu}\,_{\nu}\,\dot{x}^{\lambda} 
= \tilde{L}_{\lambda \nu}\,\dot{x}^{\lambda},
\end{equation}
hold on $x(\tau)$. For given initial data  
$x_* \in M$, $\dot{x}_* \in T_{x_*} M$, $b_* \in T^*_{x_*} M$
there exists a unique conformal geodesic $x(\tau)$,
$b(\tau)$ near $x_*$ satisfying for given $\tau_* \in \mathbb{R}$
\begin{equation}
\label{cgindat}
x(\tau_*) = x_*,\,\,\,\,\,\,\dot{x}(\tau_*) =
\dot{x}_*,\,\,\,\,\,\,b(\tau_*) = b_*.
\end{equation}
From (\ref{acgxequ}) it follows that the sign of 
$\tilde{g}(\dot{x},\dot{x})$ is preserved along $x(\tau)$, since
we have
\begin{equation}
\label{cgnormprop} 
\tilde{\nabla}_{\dot{x}}(\tilde{g}(\dot{x},\dot{x}))
= - 2 <b, \dot{x}>\,\tilde{g}(\dot{x},\dot{x}).
\end{equation}
In particular, if $\tilde{g}(\dot{x},\dot{x}) = 0$ holds at one
point it holds everywhere on $x(\tau)$ and 
(\ref{acgxequ}) implies that $x(\tau)$ can be
reparametrized to coincide with a null geodesic of
$\tilde{g}$. 

Let $x(\tau)$, $b(\tau)$ and $\bar{x}(\bar{\tau})$, 
$\bar{b}(\bar{\tau})$ be two solutions to (\ref{acgxequ}),
(\ref{bcgbequ}) with $\tilde{g}(\dot{x},\dot{x}) \neq 0$.
We derive now the conditions under which the curves
$x(\tau)$, $\bar{x}(\bar{\tau})$ coincide locally as point sets
such that there exists a local reparameterization $\tau =
\tau(\bar{\tau})$ with 
$x(\tau(\bar{\tau})) = \bar{x}(\bar{\tau})$. 
The latter relation and (\ref{acgxequ}) imply
$\dot{x}\,\partial^2_{\bar{\tau}} \tau
+ 2<\bar{b} - b, \dot{x}>\,
\dot{x}\,(\partial_{\bar{\tau}} \tau)^2
- \tilde{g}(\dot{x},\dot{x})(\bar{b} - b)
\,(\partial_{\bar{\tau}} \tau)^2 = 0$
which is equivalent to the equations 
\begin{equation}
\label{b-tauequ}
\bar{b} - b = \alpha\,\dot{x}^{\flat},\,\,\,\,\,\,
\partial^2_{\bar{\tau}} \tau + \alpha\,\tilde{g}(\dot{x},\dot{x})
\,(\partial_{\bar{\tau}} \tau)^2 = 0,
\end{equation}
with some function $\alpha$. In the first of these equations the
index of $\dot{x}$ is lowered with the metric $\tilde{g}$. 
From the first equation
and (\ref{bcgbequ}) follows
\begin{equation}
\label{alphaequ}
\dot{\alpha} = 2\,\alpha\,<b,\dot{x}> + \,\frac{1}{2}\,\alpha^2\,
\tilde{g}(\dot{x},\dot{x}),
\end{equation}
which together with (\ref{b-tauequ}) is equivalent to our requirement.
Equations (\ref{cgnormprop}), (\ref{alphaequ}) imply
$\partial_{\tau}(\alpha\,\tilde{g}(\dot{x},\dot{x}))
= 1/2\,(\alpha\,\tilde{g}(\dot{x},\dot{x}))^2$
which has the solutions 
\[
\alpha\,\tilde{g}(\dot{x},\dot{x}) 
= \frac{2\,\alpha_*\,\tilde{g}(\dot{x}_*,\dot{x}_*)}
{2 - \alpha_*\,\tilde{g}(\dot{x}_*,\dot{x}_*)\,\Delta\,\tau}, 
\]
where $\Delta\,\tau = \tau - \tau_*$ and 
$\Delta\,\bar{\tau} = \bar{\tau} - \bar{\tau}_*$. 
From these solutions follows finally with (\ref{b-tauequ})  
\begin{equation}
\label{btautraf}
\Delta\,\tau =  
\frac{4\,e\,\Delta\,\bar{\tau}}
{1 + 2\,e\,\alpha_*\,\tilde{g}(\dot{x}_*,\dot{x}_*)
\,\Delta\,\bar{\tau}},\,\,\,\,\,\,
\bar{b} =  b + \frac{1}{\tilde{g}(\dot{x},\dot{x})} 
\frac{2\,\alpha_*\,\tilde{g}(\dot{x}_*,\dot{x}_*)}
{2 - \alpha_*\,\tilde{g}(\dot{x}_*,\dot{x}_*)\,
\Delta\,\tau}\,\dot{x},
\end{equation}
\[
e, \alpha_*, \tau_*, \bar{\tau}_* \in \mathbb{R},
\,\,\,\,\,\, e \neq 0.  
\]
Thus, the changes of the initial data (\ref{cgindat})
which locally preserve the point set spread out by the curve
$x(\tau)$ are given by
\begin{equation}
\label{curvepreserve}
\dot{x}_* \rightarrow 4\,e\,\dot{x}_*,\,\,\,\,\,\,
b_* \rightarrow b_* + \alpha_*\,\dot{x},\,\,\,\,\,\,
e, \alpha_* \in \mathbb{R},\,\,\,\,\,\, e \neq 0.
\end{equation} 
The 1-form remains unchanged and an affine parameter transformation 
results if $\alpha_* = 0$. 
With suitable choices of the free constants, however, any fractional
linear transformations of the parameter can be obtained and it can
be arranged that
$\tau \rightarrow \infty$ at any prescribed value of
$\bar{\tau}$. 

This fact is related to an important difference
between conformal geodesics and metric geodesics. It may happen
that the parameter $\tau$ on a conformal geodesic takes any value
in $\mathbb{R}$ while the curve still acquires endpoints in $M$ as 
$\tau \rightarrow \pm \infty$. If the transformation in 
(\ref{btautraf}) has a pole it may further happen that the point sets
spread out by $x(\tau)$, $\bar{x}(\bar{\tau})$ coincide only
partly, namely at the points where the transformation is defined,
but each of the curves extends into a region not entered by the
other one. In fact, there may occur an arbitrary number of such
overlaps (we will encounter and make use of this phenomenon below).
In certain contexts it may be preferable to call the union of
the corresponding point sets a conformal geodesic. However,
it will be more convenient for us to preserve this name for a
solution of the conformal geodesic equation with its distinguished
parameter. 

If the 1-form $b$ is used to define a connection
$\hat{\nabla}$ along $x(\tau)$  by requiring that its difference 
tensor with respect to $\tilde{\nabla}$ is given by
\begin{equation}
\label{transpcon}
\hat{\nabla} - \tilde{\nabla} = S(b),
\end{equation} 
equation (\ref{acgxequ}) can be written in the form
$\hat{\nabla}_{\dot{x}}\dot{x} = 0$. Thus $x(\tau)$ is an
autoparallel with respect to $\hat{\nabla}$. Equation
(\ref{bcgbequ}), which determines the connection $\hat{\nabla}$ along
that autoparallel, acquires some meaning by comparing it with
(\ref{WLtens}). A congruence of conformal geodesics, covering
smoothly and without caustics an open set $U$, defines a smooth
1-form field $b$ and thus a connection $\tilde{\nabla}$ on
$U$. If $\nabla$, $f$, $L$ are replaced in (\ref{WLtens}) by
$\tilde{\nabla}$, $b$, $\tilde{L}$ respectively and transvect with
$\dot{x}$, we find that equation (\ref{bcgbequ}) takes the form of a
restriction on the connection $\hat{\nabla}$ in the direction of the
congruence, given by $\hat{L}_{\mu \nu}\,\dot{x}^{\mu} = 0$.  
Here it turns out important that (\ref{WLtens}) does not contain the 
contraction of $\nabla_{\nu}\,f_{\mu}$. 

Let $f$ be an arbitrary 1-form and  
$\hat{\nabla} = \tilde{\nabla} + S(f)$ the associated Weyl
connection.  Observing (\ref{WLtens}) with  $\nabla$, $L$ replaced by
$\tilde{\nabla}$, $\tilde{L}$, we find for any curve $x(\tau)$
in $M$ and 1-form $b(\tau)$ along $x(\tau)$ the identities
\begin{equation}
\label{hatacgxequ}
(\tilde{\nabla}_{\dot{x}}\dot{x})^{\mu}
+ S(b)_{\lambda}\,^{\mu}\,_{\rho}\,\dot{x}^{\lambda}\,\dot{x}^{\rho} 
= 
(\hat{\nabla}_{\dot{x}}\dot{x})^{\mu}
+
S(b-f)_{\lambda}\,^{\mu}\,_{\rho}\,\dot{x}^{\lambda}\,\dot{x}^{\rho}, 
\end{equation}
\begin{equation}
\label{hatbcgbequ}
(\tilde{\nabla}_{\dot{x}}b)_{\nu} - \frac{1}{2}\, 
b_{\mu}\,S(b)_{\lambda}\,^{\mu}\,_{\nu}\,\dot{x}^{\lambda} 
- \tilde{L}_{\lambda \nu}\,\dot{x}^{\lambda}
\end{equation}
\[
=
(\hat{\nabla}_{\dot{x}}(b-f))_{\nu} 
- \frac{1}{2}\, 
(b-f)_{\mu}\,S(b-f)_{\lambda}\,^{\mu}\,_{\nu}\,\dot{x}^{\lambda} 
- \hat{L}_{\lambda \nu}\,\dot{x}^{\lambda}.
\]
It follows that conformal geodesics are invariant under
transitions to general Weyl connections and, in particular, under 
conformal rescalings of $\tilde{g}$ in the
following sense: if $x(\tau)$, $b(\tau)$ is a
solution of these equations  with respect to $\tilde{\nabla}$, then
$x(\tau)$, $b(\tau) - f|_{x(\tau)}$ is a solution of the
conformal geodesic equation with respect to 
$\hat{\nabla} = \tilde{\nabla} + S(f)$. Note that the
parameter $\tau$ is an invariant of the conformal class of
$\tilde{g}$. It is determined by the initial conditions (\ref{cgindat}) 
but it does not depend on the Weyl
connection and the metric in the conformal class chosen to write the
conformal geodesics equations.

Let $e_k$, $k = 0, 1, \ldots n - 1$, be a frame field which is
parallel along $x(\tau)$ for the connection
$\hat{\nabla}$ of (\ref{transpcon}) and satisfies
$\tilde{g}(e_i, e_k) 
= \Theta^{-2}\,\,diag(1, \ldots 1, -1, \ldots, -1)$
at $x(\tau_*)$ with some $\Theta = \Theta_* > 0$. It follows then
from  (\ref{wdg}), (\ref{wpcofac}), that this relation is preserved
along $x(\tau)$ with the function $\Theta = \Theta(\tau)$
satisfying 
\begin{equation}
\label{cgcofaequ}
\tilde{\nabla}_{\dot{x}}\,\Theta = \Theta <b, \dot{x}>,\,\,\,\,\,\,
\Theta(\tau_*) = \Theta_*.
\end{equation}

Consider again a congruence of conformal geodesics as above and
initial data $\Theta_*$ chosen such that the function $\Theta$
satisfying (\ref{cgcofaequ}) is smooth and positive on $U$. 
Then the associated 1-form is given in terms of the Levi-Civita
connection $\nabla$ of the metric
$g = \Theta^2\,\tilde{g}$, for which the frame is orthonormal, by
\begin{equation}
\label{bingframe}
h(\tau) = b(\tau) - \Theta^{-1}\,d\,\Theta|_{x(\tau)},
\end{equation}
along $x(\tau)$. Equations (\ref{cgcofaequ}) and (\ref{cgnormprop}) 
imply $x(\tau)$
\begin{equation}
\label{normgauge}
<h, \dot{x}>\, = 0,\,\,\,\,\,\,
g(\dot{x}, \dot{x}) = \Theta^2\,\tilde{g}(\dot{x}, \dot{x}) =
\Theta^2_*\,\tilde{g}(\dot{x}_*, \dot{x}_*). 
\end{equation}

We assume the congruence, the function $\Theta$, and the frame $e_k$
to be constructed by a suitable choice of the initial data such
that
\begin{equation}
\label{normframeadapted}
g(\dot{x}, \dot{x}) = 1,\,\,\,\,\,\, e_0 = \dot{x}.
\end{equation}
From (\ref{normgauge}) and  
$\nabla = \tilde{\nabla} + S(\Theta^{-1}\,d\Theta)
= \hat{\nabla} - S(h)$ follows
$\nabla_{\dot{x}}e_k = 
- <h, e_k> \dot{x} + g(\dot{x}, e_k)\,h^\sharp$,
where the index of $h$ is raised with $g$. This implies that
\[
F_{\dot{x}}e_k \equiv \nabla_{\dot{x}}e_k 
- g(\dot{x}, e_k)\,\nabla_{\dot{x}}\dot{x}
+ g(\nabla_{\dot{x}}\dot{x}, e_k)\,\dot{x}
\]
\[
= - <h, e_k> \dot{x} + g(\dot{x}, e_k)\,h^\sharp
- g(\dot{x}, e_k)\,h^\sharp
+ g(h^\sharp, e_k)\,\dot{x} = 0.
\]
{\it Thus the frame $e_k$, which is parallely propagated with
respect to
$\hat{\nabla}$, is in general not parallely but always  
Fermi-propagated with respect to $\nabla$}. 

The following general observation, which follows by a direct
calculation, describes the sense in which Fermi-transport
is conformally invariant.
Let $\theta > 0$ be some conformal factor and $\tilde{g}$, $g$ 
metrics on $M$ such that $g = \theta^2\,\tilde{g}$. Denote by
$\tilde{F}$, $F$ the respective Fermi-transports.  Let
$x(\tau)$ be any curve in $M$ such that $g(\dot{x}, \dot{x}) = 1$.
For any vector field $X$ along the curve we then have
$\tilde{F}_{\theta\,\dot{x}}(\theta\,X) = 
\theta^2\,F_{\dot{x}}X$, 
where $\theta\,\dot{x}$ is the tangent vector of the
curve parametrized in terms of $\tilde{g}$-arc length. 

Applying this to the situation considered above, we see that the
$\tilde{g}$-orthonormal frame $\Theta e_k$ is Fermi-transported
along the conformal geodesics, if the latter are parametrized in
terms of $\tilde{g}$-arc length. 

Let $f: M \rightarrow M$ be a conformal diffeomorphism of
$\tilde{g}$ such that $f^*\,\tilde{g} = \Omega^2\,\tilde{g}$ with
some function $\Omega$ on $M$. Since conformal geodesics are
invariants of the conformal structure, it is not surprising that
$f$ maps conformal geodesics into conformal geodesics. If $K$ is a
conformal Killing vector field its flow
defines a 1-parameter family of local conformal diffeomorphism and
thus maps the set of conformal geodesics into itself. Though we
will occasionally make use of this fact, it will not be demonstrated
it here and we refer to \cite{ogiue} for the formal argument.

\subsection{The conformal deviation equations}

Following the discussion of the Jacobi equation for metric
geodesics, we derive now an analogous system of equations for
conformal geodesics. Let $x(\tau, \lambda)$, $b(\tau, \lambda)$ be
a family of solutions to (\ref{acgxequ}), (\ref{bcgbequ}) depending
smoothly on a parameter $\lambda$. We denote the tangent vector
field of the conformal geodesics, the {\it deviation vector field}
of the congruence, and the {\it deviation 1-form} by
\[
X = \partial_{\tau}\,x = \dot{x},\,\,\,\,\,\,
Z = \partial_{\lambda}\,x,\,\,\,\,\,\,
B = \tilde{\nabla}_Z\,b,
\]
respectively. 
Considering $x = x(\tau, \lambda)$ as a map from some open subset
of
$\mathbb{R}^2$ into
$\tilde{M}$ and denoting by $T x$ its tangent map, gives 
$Tx([\partial_{\tau}, \partial_{\lambda}]) = 0$, whence
$[X, Y] = \tilde{\nabla}_X\,Z - \tilde{\nabla}_Z\,X = 0$.
This, (\ref{acgxequ}), and 
$(\tilde{\nabla}_X\,\tilde{\nabla}_Z
- \tilde{\nabla}_Z\,\tilde{\nabla}_X)\,X 
- \tilde{\nabla}_{[X,Z]}\,Z
= \tilde{R}\,(X,\,Z)\,X$
imply the {\it conformal Jacobi
equation}
\begin{equation}
\label{xconfgeoddeviation}
\tilde{\nabla}_X\,\tilde{\nabla}_X\,Z
= \tilde{R}\,(X,\,Z)\,X - S(B; X, X) 
- 2\,S(b; X, \tilde{\nabla}_X\,Z), 
\end{equation}  
where $(S(b; X, Y))^{\mu} 
= S(b)_{\rho}\,^{\mu}\,_{\nu}\,X^{\rho}\,Y^{\nu}$
for given 1-form $b$ and vector fields $X$, $Y$.
The last two terms on the right hand side of
(\ref{xconfgeoddeviation}) indicate why conformal geodesics are
potentially more useful than metric geodesics for the construction
of coordinate systems. Under suitable circumstances
the acceleration induced by the 1-form $b$ may 
counteract curvature induced tendencies of the curves to develop
caustics.

Caustics of conformal geodesics can be more complicated than
caustics of metric geodesics because for a given tangent
vector there exists (essentially) a 3-parameter family of conformal
geodesics with the same tangent vector. Moreover, to analyse them we
need to complement the conformal Jacobi equation by   a equation
which governs the behaviour of $B$. Writing
$(\tilde{\nabla}_X\,\tilde{\nabla}_Z -
\tilde{\nabla}_Z\,\tilde{\nabla}_X)\,b  
- \tilde{\nabla}_{[X,Z]}\,b
= - b\,\tilde{R}\,(X,\,Z)$,
we get from (\ref{bcgbequ}) the {\it 1-form deviation equation}

\begin{equation}
\label{bconfgeoddeviation}
\tilde{\nabla}_X\,B =
- b\,\tilde{R}\,(X,\,Z)
+ (\tilde{\nabla}_Z \tilde{L})(X, \,.)
+ \tilde{L}(\tilde{\nabla}_X\,Z, \,.)
\end{equation}
\[
+ \frac{1}{2}\,(B \cdot S(b; X, \,.) + b \cdot S(B; X, \,.)
+ b \cdot S(b; \tilde{\nabla}_X\,Z, \,.)),
\]
where 
$(B \cdot S(b; X, \,.))_{\nu} =
B_{\mu}\,S(b)_{\rho}\,^{\mu}\,_{\nu}\,X^{\rho}$.
We refer to the equations (\ref{xconfgeoddeviation}),
(\ref{bconfgeoddeviation}) as to the system of {\it conformal
deviation equations}. They form a linear system of ODE's for
$Z$ and $B$ along the curves $x(\tau)$.

\subsection{Conformal geodesic equations on warped products}

For later applications we work out the simplifications of the
conformal geodesic equations which are obtained in the case where
$\tilde{g}$ can be written as a warped product. Thus we assume
that there exist coordinates
$x^{\mu}$ for which the set 
$I = \{0, 1, \ldots , n-1\}$ in which the index $\mu$ takes its
values can be decomposed into two non-empty subsets $I_1$, $I_2$
with $I = I_1 \cup I_2$, $\emptyset = I_1 \cap I_2$, such that in
terms of the coordinates $x^A$, $A \in I_1$, and $x^a$, 
$a \in I_2$, the metric takes the form  
\begin{equation}
\label{warpedprod}
\tilde{g} = \tilde{g}_{\mu \nu}\,dx^{\mu}\,dx^{\nu} = 
h_{AB}\,dx^A\,dx^B 
+ f^2\,k_{cd}\,dx^c\,dx^d,
\end{equation}
\[
h_{AB} = h_{AB}(x^C),\,\,\,\,\,\,
k_{ab} = k_{ab}(x^c),\,\,\,\,\,\,
1 \le p \equiv k_{ab}\,k^{ab} \le n - 1,\,\,\,\,\,\,
f = f(x^A) > 0.
\]
The Christoffel coefficients are then given by 
\[
\tilde{\Gamma}_B\,^A\,_C = \frac{1}{2}\,h^{AD}\,(h_{BD,C} + h_{CD,B} - h_{BC,D}),
\,\,\,\,\,\,\, 
\tilde{\Gamma}_b\,^A\,_c = - h^{AD}\,f\,f_{,D}\,k_{bc},
\] 
\[
\tilde{\Gamma}_B\,^A\,_c = 0,\,\,\,\,\,\,\,
\tilde{\Gamma}_B\,^a\,_C = 0,
\]
\[
\tilde{\Gamma}_B\,^a\,_c = \tilde{\Gamma}_c\,^a\,_B =
\frac{1}{f}\,f_{,B}\,k^a\,_c,\,\,\,\,\,\,\,
\tilde{\Gamma}_b\,^a\,_c = \frac{1}{2}\,k^{ad}\,(k_{bd,c} +
k_{cd,b} - k_{bc,d}).
\]  
Those which cannot be derived by symmetry
considerations from the coefficients above,
vanish. The curvature tensor has components
\begin{equation}
\label{1warpRiem}
R^A\,_{BCD}[\tilde{g}] = R^A\,_{BCD}[h],\,\,\,\,\,\,\,\,
R^a\,_{BcD}[\tilde{g}] = - k^a\,_c\,\frac{1}{f}\,D_B\,D_D\,f,
\end{equation}
\begin{equation}
\label{2warpRiem}
R^a\,_{bcd}[\tilde{g}] = R^a\,_{bcd}[k] - 2\,
h^{CD}\,f_{,C}\,f_{,D}\,k^a\,_{[c}\,k_{d]b},
\end{equation}
where $D$ denotes the $h$-Levi-Civita connection with connection
coefficients $\tilde{\Gamma}_A\,^B\,_C$. Components which cannot be
deduced by symmetry considerations from those
above, vanish. In particular
\begin{equation}
\label{vanwarpcurv}
R^A\,_{BCd}[\tilde{g}] = 0,\,\,\,\,\,\,\,
R^A\,_{bcd}[\tilde{g}] = 0,\,\,\,\,\,\,\,
R^a\,_{BCD}[\tilde{g}] = 0.
\end{equation}
This implies 
$L_{Ac}[\tilde{g}] = 0$,
\[
L_{cd}[\tilde{g}] = \frac{1}{2}\,R_{cd}[k] -
\]
\[
\frac{1}{12}\,k_{cd}\,R[k] 
- \{\,\frac{1}{12}\,R[h] + \frac{3 - p}{6}\,f\,D_AD^Af
+ \frac{(p - 1)(6 - p)}{12}\,D_AfD^Af\,\}\,k_{cd},
\]

\[
L_{AB}[\tilde{g}] = \frac{1}{2}\,R_{AB}[h] 
- \frac{1}{12}\,h_{AB}\,R[h]
- \frac{p}{2\,f}\,\{\,D_AD_Bf - \frac{1}{3}\,h_{AB}\,D_CD^Cf\,\}
\]
\[
- \frac{1}{12\,f^2}\,\{\,R[k] - p\,(p - 1)\,D_Cf\,D^Cf\,
\}\,h_{AB}.
\]
The conformal geodesic equations take the form
\begin{equation}
\label{wpxce1}
\ddot{x}^A + \tilde{\Gamma}_B\,^A\,_C\,\dot{x}^B\,\dot{x}^C + 
\tilde{\Gamma}_b\,^A\,_c\,\dot{x}^b\,\dot{x}^c =
\end{equation}
\[
- 2\,(b_C\,\dot{x}^C + b_c\,\dot{x}^c)\,\dot{x}^A
+ (h_{BC}\,\dot{x}^B\,\dot{x}^C 
+ f^2 k_{bc}\,\dot{x}^b\,\dot{x}^c)\,h^{AD}\,b_D,
\]
\begin{equation}
\label{wpxce2}
\ddot{x}^a + \tilde{\Gamma}_b\,^a\,_c\,\dot{x}^b\,\dot{x}^c +
\tilde{\Gamma}_B\,^a\,_c\,\dot{x}^B\,\dot{x}^c + 
\tilde{\Gamma}_b\,^a\,_C\,\dot{x}^b\,\dot{x}^C =
\end{equation} 
\[
- 2\,(b_C\,\dot{x}^C + b_c\,\dot{x}^c)\,\dot{x}^a
+ (h_{BC}\,\dot{x}^B\,\dot{x}^C 
+ f^2 k_{bc}\,\dot{x}^b\,\dot{x}^c)\,\frac{1}{f^2}\,k^{ac}\,b_c,
\]
\begin{equation}
\label{wpbce1}
\dot{b}_A - \tilde{\Gamma}_C\,^D\,_A\,\dot{x}^C\,b_D 
- \tilde{\Gamma}_c\,^d\,_A\,\dot{x}^c\,b_d =
\end{equation}
\[
(b_C\,\dot{x}^C + b_c\,\dot{x}^c)\,b_A
- \frac{1}{2}\,(h^{BC}\,b_B\,b_C 
+ \frac{1}{f^2}\,k^{bc}\,b_b\,b_c)\,h_{AD}\,\dot{x}^D
+ \tilde{L}_{AB}[\tilde{g}]\,\dot{x}^B,
\]
\begin{equation}
\label{wpbce2}
\dot{b}_a - \tilde{\Gamma}_c\,^D\,_a\,\dot{x}^c\,b_D 
- \tilde{\Gamma}_C\,^d\,_a\,\dot{x}^C\,b_d
- \tilde{\Gamma}_c\,^d\,_a\,\dot{x}^c\,b_d =
\end{equation}
\[
(b_C\,\dot{x}^C + b_c\,\dot{x}^c)\,b_a
- \frac{1}{2}\,(h^{BC}\,b_B\,b_C 
+ \frac{1}{f^2}\,k^{bc}\,b_b\,b_c)\,f^2\,k_{ac}\,\dot{x}^c
+ \tilde{L}_{ab}[\tilde{g}]\,\dot{x}^b.
\]

\section{Conformal geodesics on vacuum fields}

Though many of the subsequent discussions apply to more general
situations, we shall assume from now on that
$dim\,M = 4$,
$sign(\tilde{g}) = (1, -1,-1,-1)$ and that $\tilde{g}$ satisfies
Einstein's  vacuum field equations 
$\tilde{R}_{\mu \nu} = 0$.
With the resulting simplification and the notation introduced
above the conformal geodesic equations and the equation for the
frame
$e_k$ now take the form
\begin{equation}
\label{vacgxequ}
\tilde{\nabla}_{\dot{x}}\dot{x} + 2 <b, \dot{x}>\dot{x}
- \tilde{g}(\dot{x},\dot{x})\,b^{\sharp} = 0, 
\end{equation}
\begin{equation}
\label{vbcgbequ}
\tilde{\nabla}_{\dot{x}}b \,- <b,\dot{x}> b  
+ \frac{1}{2}\,\tilde{g}^{\sharp}(b, b)\,\dot{x}^{\flat} = 0,
\end{equation}
\begin{equation}
\label{vccgxequ}
\tilde{\nabla}_{\dot{x}} e_k
+ <b, \dot{x}> e_k + <b; e_k> \dot{x} 
- \tilde{g}(\dot{x}, e_k)\,b^{\sharp} = 0. 
\end{equation}
Since these equations admit solutions with vanishing 1-form $b$, 
metric geodesics form on vacuum space-times a subclass of
conformal geodesics.

\subsection{Conformal geodesics on Minkowski space}

Denote by $x^{\mu}$ coordinates on Minkowski space $(M, \eta)$ in
which the metric takes the form 
$\eta = \eta_{\mu\nu}\,dx^{\mu}\,dx^{\nu}
= (dx^0)^2 - (dx^1)^2 - (dx^2)^2 - (dx^3)^2$ and let 
$\epsilon_k = \delta^{\mu}\,_k\,\partial_{\mu}$ be the orthonormal
standard frame. There are various possibilities to determine the
solutions to (\ref{vacgxequ}), (\ref{vbcgbequ}), (\ref{vccgxequ}). The
framework of the normal conformal Cartan connection 
(cf. \cite{friedrich:AdS}), in which conformal geodesics are defined
as ``horizontal'' curves on a certain bundle allows a purely
algebraic calculation of these curves, because the bundle admits in
the case of the conformally compactified Minkowski space a purely
group theoretical description. Since the curves on the bundle space
are known explicitly there only remains the task to calculate their
projectioin onto the base space. We just give the result of the
calculation.

Besides the initial data (\ref{cgindat}) we
prescribe the value $\Theta_* > 0$ and a Lorentz transformation
$A_*$ which determines the initial value
$e_{k*} = \Theta_*^{-1}\,A^j_*\,_{k}\,\epsilon_j$ of the frame
$e_k$. The corresponding solution is given by

\begin{equation}
\label{Tmink}
\Theta(\tau) =
\Theta_*\,\left(1 + \Delta \tau\,<b_*,\dot{x}_*> 
+  \frac{1}{4}\,(\Delta \tau)^2\,\eta(\dot{x}_*,\dot{x}_*)\,
\eta^{\sharp}(b_* , b_*)\right),
\end{equation}
\begin{equation}
\label{xmink}
x^{\mu}(\tau) = x^{\mu}_* + \frac{\Theta_*}{\Theta(\tau)}\,
\left(\Delta \tau\,\dot{x}^{\mu}_* 
+ \frac{1}{2}\,
(\Delta \tau)^2\,\eta(\dot{x}_*,\dot{x}_*)\,b_*^{\mu}\right),
\end{equation}
\begin{equation}
\label{bmink}
b_{\nu}(\tau) =
(1 + \Delta \tau <b_*,\dot{x}_*>)\,b_{*\nu} - \frac{1}{2}\,\Delta
\tau\,
\eta^{\sharp}(b_*,b_*)\,\dot{x}_{*\nu},
\end{equation}
\begin{equation}
\label{ekmink}
e_k =  \frac{1}{\Theta(\tau)}\,A^j\,_k(\tau)\,\epsilon_j,
\end{equation}
with a Lorentz transformation $A(\tau) = (A^j\,_k(\tau))$ given by
\[
A(\tau) = A_* + \Delta \tau\,b_*^{\sharp}\,\dot{x}_*^{\flat} 
\]
\[
- \frac{\Theta_*}{\Theta(\tau)}\,(\Delta \tau\,\dot{x}_* +
\frac{1}{2}\,(\Delta\tau)^2\,\eta(\dot{x}_*,\dot{x}_*)\,b_*^{\sharp}
)
\,(\,b_* \cdot A_* + \frac{1}{2}\,\Delta \tau\,
\eta^{\sharp}(b_*,b_*)\,\dot{x}_*^{\flat}).
\]
Indices are moved in this subsection with the metric $\eta$.

Conformal geodesics which satisfy $\eta(\dot{x}, \dot{x}) = 0$
at one point coincide (as point sets) with null geodesics, 
those for which $b_* = 0$ are Minkowski geodesics. 
If $\dot{x}_*$ is space- or time-like we can assume, possibly after
a reparametrization, that $<b_*, \dot{x}_*>\, = 0$. It follows
that the remaining conformal geodesics fall into one of the
following classes. If $\dot{x}_*$ and $b^{\sharp}_*$ generate a
space-like 2-surface in the tangent space of $x_*$, the
corresponding conformal geodesics is a metric circle in the plane
tangent to that 2-surface. If $\dot{x}_*$ and $b^{\sharp}_*$
generate a time-like 2-surface and $\dot{x}_*$ is space-like,
the corresponding conformal geodesics is a space-like metric
hyperbola in the plane tangent to that 2-surface. If $\dot{x}_*$ is
space-like and $b_*$ is null, the conformal geodesics is a
space-like curve in a null planed.

We shall mainly be interested in the last case in which $\dot{x}_*$
and $b^{\sharp}_*$ generate a time-like 2-surface and $\dot{x}_*$
is time-like. An example of such a conformal geodesic is given by
the curve 
\[
x(\tau) = (
\frac{\tau}{1 - \frac{a^2\,\tau^2}{4}},
\,\frac{1}{a} 
+ \frac{\frac{a\,\tau^2}{2}}{1 - \frac{a^2\,\tau^2}{4}}, 
0, 0), \,\,\,\,\,\,\,\,\,\,\,\, |\tau| < \frac{2}{|a|},
\]
which is a time-like hyperbola satisfying 
$\eta_{\mu\nu}\,x^{\mu}(\tau)\,x^{\mu}(\tau) = - \,a^{-2}$.

We shall use now time-like conformal geodesics to
construct coordinate systems on Minkowski space where the parameter
$\tau$ on the curves defines a time coordinate while the other
coordinates are obtained by dragging along spatial coordinates on
$\tilde{S}$. Most important for us is the fact that the
construction provides a conformal rescaling and extension of
Minkowski space to which the coordinates are adapted in a natural
way.

We denote by $r_*$ the restriction of the standard radial coordinate
$r$ on Minkowski space to the hypersurface $\tilde{S} = \{x^0 =
0\}$. In spherical coordinates $\theta$, $\phi$ the inner metric
induced on $S$ then takes the form
$\tilde{h} = - (d r_*^2 + r_*^2\,d \sigma^2)$ with 
$d \sigma^2 = d \theta^2 + \sin^2 \theta\, d\phi^2$
the standard line element on the 2-sphere. If we write
$r_* = \tan(\frac{\chi}{2})$ with $0 \le \chi <  \pi$, we find
that the conformal factor 
$\Theta_* = 2\,(1 + r^2_*)^{-1} = 1 + \cos \chi$ allows
us to realize the conformal embedding of $\tilde{S}$ into the
3-sphere $S = \tilde{S} \cup \{i\}$ with line element
$\Theta_*^2\,\tilde{h} = - (d \chi^2 + \sin^2\chi\,d \sigma^2)$, 
where $i$ denotes the point $\{\chi = \pi\}$ at space-like infinity.

To define a congruence of conformal geodesics and a conformal
factor, we choose initial data on $\tilde{S}$ as follows. At $x_* =
(0, r_*\,u) \in \tilde{S}$, $u \in \mathbb{R}^3$ with $|u| = 1$, we set 
$b_*^{\sharp} = a\,u^c\,\epsilon_c$ (sum over $c = 1,2,3$) and
$\dot{x}_* = \Theta_*^{-1}\,\epsilon_0$ such that
$\Theta_*^2 \,\eta(\dot{x}, \dot{x}) = 1$. The function $a$ on
$\tilde{S}$ is determined by the following consideration. With the
data above and $\tau_* = 0$ we get $\Theta(\tau) = \Theta_*\,(1 -
\frac{1}{4}\,\tau^2\,a^2\,\Theta_*^{-2})$ from (\ref{Tmink}). 
To obtain, if possible, a 1-form $b$ which is exact, we require 
$\Theta^{-1}\,d\,\Theta = b$ at $x_*$ which gives
$a = 2\,r_*\,(1 + r_*^2)^{-1}$. Our data are then spherically
symmetric and we can express the conformal factor (\ref{Tmink}) and
the curves (\ref{xmink}) in terms of the time coordinate $t = x^0$ and
the radial coordinate $r$ to obtain 
\begin{equation}
\label{c1Ttrmink}
\Theta =
\Theta_*\,(1 -  \frac{1}{4}\,\tau^2\,r_*^2),\,\,\,\,\,\,
t = \frac{2\,(1 + r_*^2)\,\tau}{4 - \tau^2\,r_*^2},\,\,\,\,\,\,
r = \frac{r_*^2\,(4 + \tau^2)}{4 - \tau^2\,r_*^2}.
\end{equation}
To relate these expressions to known facts, we use 
$r_* = \tan \frac{\chi}{2}$ and set
$\frac{\tau}{2} = \tan \frac{s}{2}$ to obtain
\begin{equation}
\label{c1Tmink}
\Theta = \Omega\,\omega,
\quad\mbox{with}\quad
\Omega = \cos s + \cos \chi,\,\,\,\,\,\,\,
\omega = 1 + (\frac{\tau}{2})^2 = \frac{1}{\cos^2 \frac{s}{2}}, 
\end{equation}
\begin{equation}
\label{c1txmink}
t = \frac{\sin s}{\cos \chi + \cos s},\,\,\,\,\,\,\,
r = \frac{\sin \chi}{\cos \chi + \cos s}.
\end{equation}
Reading the last equation as a coordinate transformation, we get
\[
\Omega^2\,\eta =
\Omega^2\,(d t^2 - d r^2 - r^2\,d \sigma^2)
= g_E \equiv d s^2 - d \chi^2 - \sin^2 \chi\,d \sigma^2. 
\]
The set $M_E = \mathbb{R} \times S$ endowed with the metric $g_E$
defines the Einstein cosmos. We see that (\ref{c1txmink}) realizes the
well known conformal embedding of Minkowski space into the Einstein
cosmos, which maps the former onto the subset 
$|s \pm \chi| < \pi$, $0 \le \chi < \pi$ 
of $M_E$.
The boundary $\partial M$ of this set in $M_E$ supplies
representations of future and past null infinity, future and past
time-like infinity, and space-like infinity of Minkowski space,
which are given by the subsets
${\cal J}^{\pm} = \{s \pm \chi = \pm \pi\}$,
$i^{\pm} = \{s = \pm \pi,\,\,\,\chi = 0\}$,
$i^0 = \{s = 0,\,\,\,\chi = \pi\}$
of $\partial M$ respectively. 

In terms of $\tau$ and $\Theta$ we get from (\ref{c1Ttrmink}) 
the metric  
\begin{equation}
\label{shortextension}
\Theta^2\,\eta = \omega^2\,(
\frac{1}{\omega^2}\,d \tau^2 
- d \chi^2 - \sin^2 \chi\,d \sigma^2), 
\end{equation}
which extends smoothly to ${\cal J}^{\pm}$ but does not extend
to $i^{\pm}$ where $\omega \rightarrow \infty$.
This is not due to an unfortunate choice of intial data on
$\tilde{S}$. The curve 
$\tau \rightarrow z(\tau) 
= (s = 2\,\arctan\frac{\tau}{2}, \chi = 0)$ is
a conformal geodesic on the Einstein cosmos which
approaches $i^{\pm}$ as $\tau \rightarrow \pm \infty$. The freedom
to perform parameter transformations, characterized by
(\ref{btautraf}), (\ref{curvepreserve}), does not allow us to find a
parametrization for which curve extends simultaneously to $i^-$ and
$i^+$. Thus the separation of $i^-$ and $i^+$ realizes a
conformal invariant.  

It is possible, however, to find reparametrizations under which the
rescaled metric extends smoothly either to  $i^+$ or to $i^-$.  The
curve $\bar{\tau} \rightarrow
\bar{z}(\bar{\tau}) =  (s = s_* + 2\,\arctan (\frac{\bar{\tau}}{2}
- \tan \frac{s_*}{2}), \chi = 0)$ is again a conformal geodesic,
because $\partial_s$ is a Killing vector field for $g_E$. 
We choose $s_*$ to be a constant with $0 < s_* < \pi$. Then
$\bar{z}(\bar{\tau})$ is related to
$z(\tau)$ by the parameter transformation
$\tau = \bar{\tau}\,(1 + \tan^2 \frac{s_*}{2} -
\frac{1}{2}\,\bar{\tau}\,\tan \frac{s_*}{2})^{-1}$.
Comparing with (\ref{btautraf}) we are led to set $\tau_* = 0$,
$\bar{\tau}_* = 0$, $4\,e = (1 + \tan^2 \frac{s_*}{2})^{-1}$, and
$\alpha_*\,\eta(\dot{x}_*, \dot{x}_*) = - \tan \frac{s_*}{2}$ and
to consider the initial data  
\begin{equation}
\label{newindat}
\acute{\bar{x}}_* = \frac{1}{1 + \tan^2 \frac{s_*}{2}}\,\dot{x}_*,
\,\,\,\,\,\,
\bar{\Theta}_* =
(\eta(\acute{\bar{x}}_*,\acute{\bar{x}}_*)^{\frac{1}{2}}
= \frac{1}{\cos^2 \frac{s_*}{2}}\,\Theta_*,
\,\,\,\,\,\,
\bar{b}_* = b_* 
- \frac{\tan \frac {s_*}{2}}{\eta(\dot{x}_*, \dot{x}_*)}\,
\dot{x}_*^{\flat}.
\end{equation} 

Observing these data and setting
$\frac{\bar{\tau}}{2} = \tan \frac{s - s_*}{2} + 
\tan \frac{s_*}{2}$, we get from (\ref{xmink}) 
again equations (\ref{c1txmink}), while (\ref{Tmink}) yields now
\[
\bar{\Theta} = \bar{\Theta}_*\,(1 - \bar{\tau}\,
\frac{ \tan \frac{s_*}{2}}{1 + \tan^2 \frac{s_*}{2}}
+ \frac{1}{4}\,\bar{\tau}^2\,\frac{\tan^2 \frac{s_*}{2} 
- \tan^2   \frac{\chi}{2}}
{(1 + \tan^2 \frac{s_*}{2})^2})
= \Omega\,\,\bar{\omega},
\]
with $\Omega$ as in (\ref{c1Tmink}) and
$\bar{\omega} = \cos^{-2} \frac{(s - s_*)}{2}$.
It follows that the metric 
\[
\bar{\Theta}^2\,\eta =
\bar{\omega}^2\,(d s^2 - d \chi^2 - \chi^2\,d\,\sigma^2)
= \bar{\omega}^2\,(\frac{1}{\bar{\omega}^2}\,d\bar{\tau}^2 
- d \chi^2 - \chi^2\,d\,\sigma^2),
\]
extends regularly onto the domain $|s - s_*| < \pi$ containing
$i^+$. 

If we insist on a conformal factor $\Theta$ which
defines a conformal compactification of $(\tilde{S}, \tilde{h})$
and initial data ensuring $b = \Theta^{-1}\,d\,\Theta$
on $\tilde{S}$, invariance under $t \rightarrow - \,t$ implies the
existence of a point on $\tilde{S}$ where $b$ vanishes. The
conformal geodesic through such a point will then be a metric
geodesic. In our first example such a point is given by $r_* = 0$
and it follows from (\ref{c1Ttrmink}) that the parameter $\tau$ takes
values in $\mathbb{R}$ and the conformal factor is constant on the curve
$r = 0$. Thus, in order to achieve a compactification which extends
smoothly to $i^+$ for a finite value of the parameter we have to
choose non-time-symmetric initial data such as (\ref{newindat}).

Notice that $s_*$ could have been chosen to be a function on 
$\tilde{S}$. The possibility to change the parametrization while
leaving the curves as point sets unchanged offers a large freedom
to select slices of constant parameter value.

\subsection{The conformal factor and the 1-form on general\\ 
vacuum fields}

In the following we shall derive some
general, explicit information on the solutions to
(\ref{vacgxequ}), (\ref{vbcgbequ}), (\ref{vccgxequ}) (cf. the more general
discussion in \cite{friedrich:AdS} on solutions to the vacuum field
equations
$\tilde{R}_{\mu \nu} = \lambda\,\tilde{g}_{\mu \nu}$ with a
cosmological constant $\lambda$).

From (\ref{vacgxequ}), (\ref{vbcgbequ}) follows
\begin{equation}
\label{bxdotequ}
\tilde{\nabla}_{\dot{x}} <b, \dot{x}>\, = - <b, \dot{x}>^2 
+ \frac{1}{2}\,\tilde{g}(\dot{x},\dot{x})\,\tilde{g}^{\sharp}(b, b), 
\end{equation}
\begin{equation}
\label{bsquequ}
\tilde{\nabla}_{\dot{x}}\,\tilde{g}^{\sharp}(b, b)
= \,<b, \dot{x}>\,\tilde{g}^{\sharp}(b, b), 
\end{equation}
where the index {\it sharp} on the symbol of a metric indicates here
and in the following the contravariant version of that metric.

Assume again that the congruence of conformal geodesics,
$\Theta$, and the frame have been chosen such that
(\ref{normframeadapted}), whence 
$\tilde{g}(\dot{x}, \dot{x}) = \Theta^{-2}$
holds along the congruence.  Equation
(\ref{cgnormprop}) then implies
$\tilde{\nabla}_{\dot{x}}\,\Theta  = \Theta<b,\dot{x}>$. 
Taking a
derivative and observing (\ref{bxdotequ}) yields  
$\tilde{\nabla}^2_{\dot{x}}\,\Theta
= 1/2\,
\tilde{g}^{\sharp}(b, b)\,\Theta^{-1}$.
Taking another derivative and
observing (\ref{bsquequ}) gives finally 
\begin{equation}
\label{d3-1n}
\tilde{\nabla}_{\dot{x}}^3\,\Theta = 0.
\end{equation}
In terms of the initial data (\ref{cgindat}) this yields with  
$\Theta_* = \sqrt{(\tilde{g}(\dot{x}_*, \dot{x}_*))}$ 
the explicit expression
\begin{equation}
\label{Theta}
\Theta(\tau) = \Theta_* + \Delta \tau\,\dot{\Theta}_* 
+ \frac{1}{2}\,(\Delta \tau)^2\,\ddot{\Theta}_*
\end{equation}
\[
= \Theta_*\,\left(1 + \Delta \tau\,<b_*,\dot{x}_*>
+ \frac{1}{4}\,(\Delta \tau)^2\,\tilde{g}(\dot{x}_*, \dot{x}_*)\,
\tilde{g}^{\sharp}(b_*, b_*)\right).
\]
Similar steps lead to 
$\tilde{\nabla}_{\dot{x}}\,(\tilde{g}(\dot{x},\dot{x})\,
(\tilde{g}^{\sharp}(b, b))^2) = 0$, whence
\begin{equation}
\label{dn2b4}
\Theta^{-1}\,\tilde{g}^{\sharp}(b, b) = 
\Theta_*^{-1}\,\tilde{g}^{\sharp}(b_*, b_*) = 2\,\ddot{\Theta}_*.
\end{equation}
In particular, the sign of $\tilde{g}^{\sharp}(b,
b)$ is preserved as long as $\Theta > 0$.
From (\ref{vbcgbequ}), (\ref{vccgxequ}) follows
$\tilde{\nabla}_{\dot{x}}(\Theta<b, e_k>) = 1/2\,\Theta\,
\tilde{g}^{\sharp}(b, b)\,\tilde{g}(\dot{x},e_k)$.
From this equation and (\ref{cgcofaequ}), (\ref{normframeadapted}) 
the following explicit expression can be derived for the components 
$b_k \equiv \,<b, e_k>$ of the 1-form
$b$ in the frame $e_k$
\begin{equation}
\label{bcom}
b_0 = \Theta^{-1} \dot{\Theta},\,\,\,\,\,\,
b_a = \Theta^{-1} d_a 
\quad\mbox{with}\quad
d_a = \,<b_*, \Theta_*\, e_{k*}>.
\end{equation} 
This implies with (\ref{dn2b4}) the relation
\begin{equation}
\label{surprel}
\dot{\Theta}^2 - \delta^{ab}\,d_a\,d_b
= \Theta^2\,\eta^{jk}\,b_j\,b_k
= \Theta^2\,g^{\sharp}(b, b) 
= \tilde{g}^{\sharp}(b, b)
= \Theta\,\,\Theta_*^{-1}\,\tilde{g}^{\sharp}(b_*, b_*)
= 2\,\Theta\,\ddot{\Theta}_*.
\end{equation}
While the expressions (\ref{Theta}), (\ref{bcom}) depend on invariants
built from the initial data, they are universal in the sense that
their form does not depend on the solution $\tilde{g}$.

Using the representation of the 1-form with respect to the
connection of $g$, i.e.  (\ref{bingframe}), we can write by (\ref{bcom})
on the congruence
$e_k(\Theta) = \Theta<b, e_k> - \,\Theta<h, e_k>$. 
If there is a point on a given curve of the congruence where $\Theta$
vanishes and $\Theta\,b$ and $h$ remain bounded on the curve, 
(\ref{bcom}), (\ref{surprel}) imply
\begin{equation}
\label{scrinull}
\eta^{jk}e_j(\Theta)\,e_k(\Theta) \rightarrow 0 
\quad\mbox{as}\quad \Theta \rightarrow 0.
\end{equation}

\subsection{The $\tilde{g}$-adapted form of the conformal geodesic\\
equation}

We write $b = \hat{b} + \zeta\,\dot{x}^{\flat}$ (where here and
below  indices of 1-forms and vectors are moved with $\tilde{g}$)
with
$\hat{b}$ such that
\begin{equation}
\label{bsplit}
<\hat{b}, \dot{x}>\, = 0, \quad\mbox{whence}\quad
\zeta = \frac{<b, \dot{x}>}{\tilde{g}(\dot{x}, \dot{x})},\,\,\,\,\,\,
g^{\sharp}(b, b) = \,<b,\dot{x}>^2 +\,
\, g^{\sharp}(\hat{b}, \hat{b}). 
\end{equation}
Equations (\ref{vacgxequ}), (\ref{vbcgbequ}) are then
equivalent to
$\tilde{\nabla}_{\Theta\,\dot{x}}\,\Theta\,\dot{x} =
\hat{b}^{\sharp}$,
$\tilde{\nabla}_{\Theta\,\dot{x}}\,\hat{b} 
= - \tilde{g}^{\sharp}(\hat{b},\hat{b})\,
\Theta\,\dot{x}^{\flat}$. We introduce the parameter transformation
\begin{equation}
\label{partraf}
\bar{\tau}(\tau) = \bar{\tau}_* 
+ \int_{\tau_*}^{\tau} \frac{d\tau'}{\Theta(\tau')}, 
\end{equation} 
with inverse $\tau = \tau(\bar{\tau})$
and set $\bar{x}(\bar{\tau}) = x(\tau(\bar{\tau}))$. Then
$\bar{x}' \equiv \partial_{\bar{\tau}}\bar{x}
= \Theta\,\dot{x}$ satisfies
$\tilde{g}(\bar{x}',\bar{x}') = 1$ and we obtain the
{\it $\tilde{g}$-adapted conformal geodesic equations} 
\begin{equation}
\label{cgvacadapted}
\tilde{\nabla}_{\bar{x}'}\,\bar{x}' = \hat{b}^{\sharp},
\,\,\,\,\,\,\,
\tilde{\nabla}_{\bar{x}'}\,\hat{b} = \beta^2\,\bar{x}^{'\flat},
\end{equation}
where, by (\ref{bcom}), 
\begin{equation}
\label{b'2preserved}
\beta^2 \equiv
- \tilde{g}^{\sharp}(\hat{b},\hat{b}) = 
\delta^{ab}\,d_a\,d_b =
const. 
\end{equation}
along
the conformal geodesic.
These equations bring out the important role of $\hat{b}$. If
$\hat{b}$ vanishes at a point it vanishes along
$\bar{x}(\bar{\tau})$ and the curve is a $\tilde{g}$-geodesics.
 
We determine the transformation (\ref{partraf}) under the assumption
that
\begin{equation}
\label{betapos}
g^{\sharp}(b_*, b_*)
= \tilde{g}(\dot{x}_*, \dot{x}_*)\,
\tilde{g}^{\sharp}(b_*, b_*) < 0.
\end{equation}
It follows then from (\ref{Theta}) that $\Theta(\tau)$ vanishes at
\begin{equation}
\label{Thetazeros}
\tau_{\pm} =  \tau_* 
- \frac{2\,\Theta_*}
{\Theta_*<b_*, \dot{x}_*> \mp \,|\beta|},
\end{equation}
with $\tau_+ \neq \tau_-$, such that
\begin{equation}
\label{Thetaoftau} 
\Theta = \frac{1}{4}\,\Theta_*\,g^{\sharp}(b_*, b_*)
(\tau - \tau_+)\,(\tau - \tau_-),
\end{equation}
and the integration of (\ref{partraf}) gives
\begin{equation}
\label{taubaroftau}
\bar{\tau}(\tau) = \bar{\tau}_*
+ \frac{1}{|\beta|}
\,\log \frac{(\tau_* - \tau_+)\,(\tau - \tau_-)}
{(\tau - \tau_+)\,(\tau_* - \tau_-)},
\end{equation}
whence
\begin{equation}
\label{tauoftaubar}
\Delta \tau = 
\frac{2\,\Theta_*\,\sinh (\frac{|\beta|}{2}\,\Delta \bar{\tau})} 
{|\beta|\,
\cosh (\frac{|\beta|}{2}\,\Delta \bar{\tau}) 
\,\,- \,\Theta_*<b_*, \dot{x}_*>\,
\sinh (\frac{|\beta|}{2}\,\Delta \bar{\tau})}.
\end{equation}
In terms of the parameter $\bar{\tau}$ we also get
\begin{equation}
\label{Thetaoftaubar}
\Theta = 
\frac{\Theta_*\,\beta^2} 
{\left(|\beta|\,
\cosh (\frac{|\beta|}{2}\,\Delta \bar{\tau}) 
\,\,- \Theta_*<b_*, \dot{x}_*>\,
\sinh (\frac{|\beta|}{2}\,\Delta \bar{\tau})\right)^2}.
\end{equation}

Suppose that the solution admits a smooth conformal extension for
which $\Theta = 0$ on ${\cal J}^+$, that the extension 
admits a point $i^+$ such that ${\cal J}^+$ coincides with the past
light cone of $i^+$, and that one of our conformal geodesics,
$x(\tau)$ say, passes through $i^+$ for a finite value $\tau_i$ of
$\tau$. Then condition
(\ref{betapos}) precludes a dicussion of the zero of $\Theta$ on
$x(\tau)$. If (\ref{Thetazeros}) describes the zeros of $\Theta$
on the conformal geodesics covering a neighbourhood of $x(\tau)$,
then, approaching $x(\tau)$, we should find $\tau_{\pm} \rightarrow
\tau_i$, and consequently
$\beta = 0$, $<b_*, \dot{x}_*>\,\neq 0$ on $x(\tau)$ and
$\tau_i =  \tau_* - \frac{2}{<b_*, \dot{x}_*>}$.

\subsection{The $\tilde{g}$-adapted conformal deviation
equations}

Denote by 
$\bar{x}(\bar{\tau}, \lambda)$, $\hat{b}(\bar{\tau}, \lambda)$ a
smooth family of  solutions to (\ref{cgvacadapted}) with family
parameter $\lambda$. As before, we write
\begin{equation}
\label{tangconnvect}
X = \partial_{\bar{\tau}}\,\bar{x} = \bar{x}',\,\,\,\,\,\,
Z = \partial_{\lambda}\,\bar{x},\,\,\,\,\,\,
\hat{B} = \tilde{\nabla}_Z\,\hat{b}.
\end{equation}
Following the derivation of the conformal deviation equations and
observing that vacuum field equations $\tilde{L}_{\mu \nu} = 0$, we
obtain the {\it $\tilde{g}$-adapted conformal deviation equations}
\begin{equation}
\label{vacadaptxconfdeviation}
\tilde{\nabla}_X\,\tilde{\nabla}_X\,Z
= C\,(X,\,Z)\,X + \hat{B}^{\sharp}.
\end{equation}  
\begin{equation}
\label{vacadaptb'confdeviation}
\tilde{\nabla}_X\,\hat{B}
= - \hat{b}\,C\,(X,\,Z)
+ (\tilde{\nabla}_Z\,\tilde{g}^{\sharp}(\hat{b},
\hat{b}))\,X^{\flat} 
+ \tilde{g}^{\sharp}(\hat{b}, \hat{b})
\,\tilde{\nabla}_X\,Z^{\flat},
\end{equation}
where $C$ denotes the conformal Weyl tensor of $\tilde{g}$.

While in general   
$\tilde{\nabla}_Z(\tilde{g}^{\sharp}(\hat{b},\hat{b})) \neq 0$,
we have by (\ref{b'2preserved}) always 
$\tilde{g}^{\sharp}(\hat{b}, \hat{b}) = const.$ along
$\bar{x}(\bar{\tau})$, which implies
$\tilde{\nabla}_X\,
(\tilde{\nabla}_Z\tilde{g}^{\sharp}(\hat{b},\hat{b})) = 
\tilde{\nabla}_Z\,(\tilde{\nabla}_X\tilde{g}^{\sharp}
(\hat{b},\hat{b})) = 0$.
Thus the coefficients of $X^{\flat}$ and 
$\tilde{\nabla}_X\,Z^{\flat}$ in the second line of 
\ref{vacadaptb'confdeviation} are constant and known along 
$\bar{x}(\bar{\tau})$ by their initial data at some $\bar{\tau}_*$.

\subsection{Conformal geodesics and the $\tilde{g}$-adapted 
conformal\\ 
deviation equations on warped product vacuum fields}

It will be shown now that on warped product vacuum fields the
discussion of the conformal geodesic equations reduces under suitable
assumptions to the analysis of one equation and that a similar
statement is true for the $\tilde{g}$-adapted conformal deviation
equations.

If $\tilde{g}$ can be written in the form (\ref{warpedprod}), the
vacuum field equations take the form
\begin{equation}
\label{warpvacequ}
R_{AC}[h] = \frac{p}{f}\,D_AD_Cf,\,\,\,\,\,\,\,\,
R_{ac}[k] = f^{- (p - 2)}\,D_A(f^{(p - 1)}\,D^Af)\,k_{ac},
\end{equation}
and the dependence of the various fields in the second equation on
$x^a$ implies
\begin{equation}
\label{conswarpvacequ}
R_{ac}[k] = \frac{R[k]}{p}\,k_{ac},\,\,\,\,\,\,\,\,
R[k] = p\,f^{- (p - 2)}\,D_A(f^{(p - 1)}\,D^Af) = const.
\end{equation}

We shall assume from now on that $p = 2$ and that the indices $A$,
$a$ take values $0, 1$ and $2, 3$ respectively. Then 
$R_{ABCD}[h] = R[h]\,h_{A[C}\,h_{D]B}$ and
$R_{AC}[h] = \frac{1}{2}\,R[h]\,h_{AC}$ and thus
$2\,D_AD_Bf = h_{AB}\,D_CD^Cf$ by the first of
equations (\ref{warpvacequ}). Contracting with $D^A$, commuting
derivatives yields $D_B(f^2\,D_AD^Af) = 0$
if (\ref{warpvacequ}) is used again. It follows 
\begin{equation}
\label{intcdef}
2\,c \equiv f^2\,D_AD^Af = const.,\,\,\,\,\,\,
D_AD_Bf = \frac{c}{f^2}\,h_{AB},\,\,\,\,\,\,
R_{ABCD}[h] = \frac{4\,c}{f^3}\,h_{A[C}\,h_{D]B}.
\end{equation}
These equations imply with (\ref{1warpRiem}), (\ref{2warpRiem}) 
\begin{equation}
\label{vacwarpRiem}
R^A\,_{BCD}[\tilde{g}] =
\frac{4\,c}{f^3}\,h^A\,_{[C}\,h_{D]B},\,\,\,
R^a\,_{bcd}[\tilde{g}] = \frac{4\,c}{f}\,k^a\,_{[c}\,k_{d]b}
,\,\,\,
R^a\,_{BcD}[\tilde{g}] = - \frac{c}{f^3}\,k^a\,_c\,h_{BD}.
\end{equation}
It follows that 
$R^{\mu}\,_{\nu \lambda \rho} = 0$ unless $c \neq 0$.

If Einstein's vacuum field equation $\tilde{L}_{\mu\nu} = 0$ holds,
equations 
(\ref{wpxce1}), (\ref{wpxce2}), (\ref{wpbce1}), (\ref{wpbce2})
admit solutions satisfying 
$\dot{x}^a = 0$, $b_c = 0$,
and these solutions obey equations which can be written in the
form  
\begin{equation}
\label{2redcgs}
D_{\dot{x}}\dot{x} = - 2<b, \dot{x}>\dot{x} + h(\dot{x},
\dot{x})\,b^{\sharp},\,\,\,\,\,\,\,\,
D_{\dot{x}}b =\,\,<b, \dot{x}>b 
- \frac{1}{2}\,h^{\sharp}(b, b)\,\dot{x}^{\flat},
\end{equation}
where all quantities are derived from $h$. We shall discuss now the
$\tilde{g}$-adapted version of
(\ref{2redcgs}), assuming that $\Delta \equiv \det(h_{AB}) < 0$.
With
\begin{equation}
\label{2heps}
\epsilon_h = \sqrt{|\Delta|}\,d\,x^0 \wedge d\,x^1,
\end{equation}
it follows that
$\epsilon_h(\dot{x},\,.) =
\sqrt{|\Delta|}(\dot{x}^0\, d\,x^1 - \dot{x}^0\,d\,x^0)$ and
$h^{\sharp}(\epsilon_h(\dot{x},\,.), \epsilon_h(\dot{x},\,.)) = 
- h(\dot{x}, \dot{x})$.
 Since the vectors $\dot{x}$, $\hat{b}^{\sharp}$ are contained in
the 2-dimensional space spanned by $\partial_A$, $A = 0, 1$, and
$<\hat{b}, \dot{x}>\, = 0$, the equations above and
(\ref{b'2preserved}) imply the representation
\begin{equation}
\label{b'repr}
\hat{b} = \pm\,\beta\,\epsilon_h(\Theta\,\dot{x},\,.)
= \pm\,\beta\,\epsilon_h(\bar{x}',\,.),
\end{equation}
where the sign is determined by the initial conditions and the
convention for $\beta$. Using this expression in the second of
equations
\ref{cgvacadapted}, gives
$\beta^2\,\bar{x}' = D_{\bar{x}'}\hat{b} =
\pm \,\beta\,\epsilon_h(D_{\bar{x}'}\,\bar{x}',\,.)$, 
which can be rewritten in the form
\begin{equation}
\label{redcgvacadapted}
D_{\bar{x}'}\,\bar{x}' = \hat{b}^{\sharp}.
\end{equation}
Thus, the $\tilde{g}$-adapted forms of the two
equations (\ref{2redcgs}) are equivalent to each other 
and we only have to solve (\ref{redcgvacadapted}) with $\hat{b}$ 
as in (\ref{b'repr}) to obtain a solution to (\ref{2redcgs}).

Observing the formulae for the connection coefficients, the
results (\ref{vanwarpcurv}) for the curvature tensor of warped
products, and
\begin{equation}
\label{projWeyl}
C^A\,_{BCD}[\tilde{g}] =
\frac{4\,c}{f^3}\,h^A\,_{[C}\,h_{D]B},
\end{equation}
equations (\ref{vacadaptxconfdeviation}), 
(\ref{vacadaptb'confdeviation}) are seen to be equivalent to each
other and to the equation
\begin{equation}
\label{2vacadaptxconfdeviation}
D_X\,D_X\,Z
= \frac{2\,c}{f^3}\,\epsilon_h(X, Z)\,\epsilon_h(X,\,.)^{\sharp} 
\pm \left(D_Z\,\beta\,\epsilon_h(X,\,.)^{\sharp}
+ \beta\,\epsilon_h(D_X\,Z,\,.)^{\sharp}\right).
\end{equation}
The sign here is chosen as in (\ref{b'repr}) and use has been made 
of the identities
\[
\epsilon_h(\epsilon_h(X,\,.)^{\sharp}\,.) = X^{\flat},
\,\,\,\,\,\,\,\,
X\,h(Z, X) - Z = \epsilon_h(X, Z)\,\epsilon_h(X,\,.)^{\sharp}.
\]

\section{Conformal geodesics on the Schwarzschild\\ 
space-time}

For our discussions various forms of the Schwarzschild
line element will be needed. Its standard form with mass $m$ is given
for 
$\bar{r} > 2\,m$ by
\begin{equation}
\label{standSchwarz}
\tilde{g} = \left(1 - \frac{2\,m}{\bar{r}}\right)\,d\,t^2
-  \left(1 - \frac{2\,m}{\bar{r}}\right)^{-1}\,d\,\bar{r}^2
- \bar{r}^2\,d\,\sigma^2.
\end{equation} 

In terms of the retarded and advanced null
coordinates
\begin{equation}
\label{retadvnull}
w = t - (\bar{r} + 2\,m\,\log(\bar{r} - 2\,m)),\,\,\,\,\,\,\,
v = t + (\bar{r} + 2\,m\,\log(\bar{r} - 2\,m)),
\end{equation}
the line element (\ref{standSchwarz}) is obtained in the forms
\begin{equation}
\label{retadvnullSchw}
\tilde{g} = \left(1 - \frac{2\,m}{\bar{r}}\right) d w^2 
+ 2\,dw\,d\bar{r} - \bar{r}^2 d \sigma^2,\,\,\,
\tilde{g} = \left(1 - \frac{2\,m}{\bar{r}}\right) d v^2 
- 2\,dv d \bar{r} - \bar{r}^2 d \sigma^2,
\end{equation}
respectively. These extend analytically into regions where 
$\bar{r} \le 2\,m$. The retarded null coordinate $w$ extends smoothly
through ${\cal J}^+$ and through the past horizon to the past
singularity, while the advanced null coordinate $v$ extends smoothly
through ${\cal J}^-$ and the future horizon to the future
singularity. 

The transformation
\[
s = \left(\frac{\bar{r}}{2\,m} - 1 \right)^{1/2} 
\,\exp (\frac{\bar{r}}{4\,m})
\,\sinh (\frac{t}{4\,m}) 
,\,\,\,\,\,\,\,
\rho = \left(\frac{\bar{r}}{2\,m} - 1 \right)^{1/2} 
\,\exp (\frac{\bar{r}}{4\,m})
\,\cosh (\frac{t}{4\,m}),
\]
applied to (\ref{standSchwarz}) with $m > 0$ produces the
Schwarzschild-Kruskal line element
\begin{equation}
\label{SchwKrusk}
\tilde{g} = K\,(d\,s^2 - d\,\rho^2) - \bar{r}^2\,d\,\sigma^2
\quad\mbox{with}\quad
K = \frac{32\,m^3}{\bar{r}}\,\exp(- \frac{\bar{r}}{2m}),
\,\,\,\,\,\,\,\,\,
\bar{r} = k(\rho^2 - s^2),
\end{equation}
where $k$ denotes the inverse of the map
$]0, \infty[ \,\, \ni \bar{r} \longrightarrow 
(\frac{\bar{r}}{2\,m} - 1)\,\exp(\frac{\bar{r}}{2m}) \in \,\,
]-1, \infty[$.
This line element extends analytically to a
non-degenerate line element on the domain $\rho^2 - s^2 > - 1$.

The coordinate $r$, given for $\bar{r} > 2\,m$
by 
\begin{equation}
\label{Schwarz-isotropic}
\bar{r} = \frac{1}{r}\,\left(r + \frac{m}{2}\right)^2
\quad\mbox{resp.}\quad
r = \frac{1}{2}\,
\left(\bar{r} - m + \sqrt{\bar{r}(\bar{r} - 2\,m)}\right),
\end{equation}
yields the isotropic Schwarzschild line element
\begin{equation}
\label{isotrSchw}
\tilde{g} = 
\left(\frac{1 - \frac{m}{2\,r}}{1 + \frac{m}{2\,r}}\right)^2 d\,t^2
- (1 + \frac{m}{2\,r})^4 (d\,r^2 + r^2 d\,\sigma^2),
\,\,\,\,\,\,\,\,\,r > \frac{m}{2}.
\end{equation}
On the hypersurface $\{t = 0\}$ it induces initial data which can be
extended analytically to an initial data set
\begin{equation}
\label{SchwKdat}
(\tilde{S} = \mathbb{R}^3 \setminus \{0\},\,\,\tilde{h} 
= - (1 + \frac{m}{2\,r})^4\,( d\,r^2 + r^2\,d\,\sigma^2),
\,\,\tilde{\chi} = 0),
\end{equation}
where $\tilde{\chi}$ denotes the second fundamental form. These data
may be identified isometrically with the initial data induced by the
Schwarzschild-Kruskal metric \ref{SchwKrusk} on the hypersurface 
$\{s = 0\}$ by the transformation
\[
\rho = \frac{r - \frac{m}{2}}{\sqrt{2\,m\,r}}\,
\exp(\frac{(r + \frac{m}{2})^2}{4\,m\,r}).
\]
In the following discussions one may think of the conformal geodesics as
being constructed on the Schwarzschild-Kerr solution (\ref{SchwKrusk})
with initial data being prescribed on the hypersurface 
$\tilde{S} = \{s = 0\}$. It will be convenient, however, to specify them
in terms of the coordinate $r$ in (\ref{SchwKdat}). To discuss the
different aspects of these curves we will use those coordinates which
will appear to give the simplest formulae. Because of the symmetries $s
\rightarrow - s$ and 
$\rho \rightarrow - \rho$, it will be sufficient to consider the  region
$\{s \ge 0, \rho \ge  0\}$ if we prescibe data for the conformal 
geodesics which respect these symmetries. The data will also be
spherically symmetric.

The
function $\bar{r}$ extends by the first of equations 
(\ref{Schwarz-isotropic}) to an analytic function on 
$\tilde{S}$ which takes its minimum $\bar{r}_{\min} = 2\,m$
at the ``throat'' $\{ r = r_h \equiv \frac{m}{2} \}$. The space 
$(\tilde{S}, \tilde{h})$ has two asymptotically flat ends.
In terms of (\ref{SchwKdat}) the end $i_1$ at
$r = \infty$ has the ``usual'' representation, while the end $i_2$ at 
$r = 0$ has a coordinate representation adapted to the metric 
$(1 + \frac{m}{2\,r})^{-4}\,\tilde{h}$ which may be
thought of as a conformal compactification of $\tilde{h}$ at the end
$i_2$. The map 
\begin{equation}
\label{Schwreflisom}
r \rightarrow (\frac{m}{2})^2\,\frac{1}{r},
\end{equation}
which corresponds to the isometry $\rho \rightarrow - \rho$ allows us to
show the equivalence of the spatial infinities $i_1$ and $i_2$. It
leaves $\bar{r}$ invariant and has $\{ r = r_h \}$ as its fixed point
set.

\subsection{Conformally compactified Schwarzschild-Kruskal data
and initial data for a congruence of conformal geodesics}

We shall in the following prescribe initial data on $\tilde{S}$ for a
congruence of conformal geodesics whose $\tilde{g}$-adapted initial
vectors $\bar{x}'$ coincide with the future directed unit normals of
$\tilde{S}$. By $\bar{r}_*$ will be denoted the restriction to
$\tilde{S}$ of the function $\bar{r}$ on the Schwarzschild-Kruskal
space-time and $r$ will be considered as a coordinate on $\tilde{S}$. For
the conformal factor we choose
\begin{equation}
\label{2Theta}
\Theta_* = \frac{1}{\bar{r}_*^2} 
= \frac{r^2}{(r + \frac{m}{2})^{4}}
\end{equation} 
which implies
\begin{equation}
\label{2rescdata}
- \Theta_*^2\,\tilde{h} = 
(r  +  \frac{m}{2})^{-4}\,(d\,r^2 + r^2\,d\,\sigma^2).
\end{equation}
It follows that the transformation 
$r = \tan \frac{\chi}{2}$, $0 < \chi < \pi$ (and, to get simple
equations, $\frac{m}{2} = \tan \frac{\mu}{2}$) could be used to make the
conformal compactification achieved by $\Theta_*$ manifest in terms of
coordinates, since it realizes an embedding of $\tilde{S}$ into  the
unit 3-sphere $S^3$ with the poles at $\chi = 0$, $\chi = \pi$
corresponding to the ends $i_2$, $i_1$ respectively.

The choice (\ref{2Theta}) is particularly well adapted to the
Schwarzschild-Kruskal geometry and implies simple formulae. However, 
it is important to note that our basic results do not depend on it. For
the analysis of the field near space-like infinity other choices might be
preferable (cf. \cite{friedrich:i0}), because the metric (\ref{2rescdata})
does not extend smoothly to space-like infinity.

We set furthermore
\begin{equation}
\label{2bdata}
b_* = \hat{b}_* = \Theta_*^{-1}\,d\,\Theta_* =
- \frac{2\,(r - \frac{m}{2})}{r\,(r + \frac{m}{2})}\,d\,r, 
\end{equation}
and, observing
$F \equiv 1 - 2\,m/\bar{r}_* = 
(r - \frac{m}{2})^2\,(r + \frac{m}{2})^{-2}$,
choose
\begin{equation}
\label{2beta}
\beta =
\frac{2\,r\,(r - \frac{m}{2})}{(r + \frac{m}{2})^3},
\end{equation}
such that $\beta$ is analytic on $\tilde{S}$ and 
$\beta = \sqrt{- \tilde{g}^{\sharp}(\hat{b}_*, \hat{b}_*)}
= \frac{2}{\bar{r}_*}\,\sqrt{F(\bar{r}_*)} > 0$
near the end $i_1$ in analogy to our procedure on Minkowski space.
If $\tilde{h}$ is used to raise indices, $\hat{b}_*^{\sharp}$ is
outward pointing at the ends $i_1$, $i_2$, as it should.
With the choices above the general formula (\ref{Thetaoftau}) reduces to
\begin{equation}
\label{2Thetaoftau}
\Theta = F(\bar{r}_*)\,
\left((\frac{2\,\Theta_*}{\beta})^2 - \tau^2\right),
\end{equation}
where functions with subscript $*$ are assumed to be constant along the
conformal geodesics.
Note that our initial data and gauge conditions are
preserved by the isometry (\ref{Schwreflisom}), and are adapted to
the spherical symmetry and the time reflection symmetry. We have $\beta
\rightarrow - \beta$ under (\ref{Schwreflisom}) and $\beta(r) = 0$
precisely for $r = \frac{m}{2}$. This means
that  conformal geodesic satisfying the initial data above at
points with $r = \frac{m}{2}$ will be metric geodesics in the
hypersurface $\{\rho = 0\}$ of the Schwarzschild-Kruskal space-time.

\subsection{The conformal geodesic equations on the Schwarzschild
space-time}

To discuss conformal geodesics on the Schwarzschild space-time we
write the $\tilde{g}$-adapted form of the conformal geodesic equation
in terms of the line element
(\ref{standSchwarz}), which can be written in the form 
(\ref{warpedprod}) 
with the metric $h$ given by
\[
h = F\,dt^2 - \frac{1}{F}\,d\bar{r}^2,\,\,\,\,\,\,\,\,\,
F = (1 - \frac{2\,m}{\bar{r}}).
\]
Initial data for the conformal geodesics will be given on the
hypersurface $\tilde{S} = \{t = 0\}$ with $\tau_* = 0$,
$\bar{\tau}_* = 0$ on $\tilde{S}$.

Because of (\ref{b'repr}) (where due to our conventions we have to use
the minus sign), equations (\ref{redcgvacadapted}) take the form
\begin{equation}
\label{1xcgschw}
t'' + \frac{F,_{\bar{r}}}{F}\,\bar{r}'\,t' = 
\frac{1}{F}\,\beta\,\bar{r}',
\end{equation}
\begin{equation}
\label{2xcgschw}
\bar{r}'' - \frac{F,_{\bar{r}}}{2\,F}\,(\bar{r}')^2
+ \frac{F\,F,_{\bar{r}}}{2}\,(t')^2 =
F\,\beta\,t', 
\end{equation}
with $\beta$ satisfying (\ref{b'2preserved}).
Assuming on $\tilde{S}$ that the initial vector is the future directed
unit normal to $\tilde{S}$, the initial data on $\tilde{S}$ are given
by
\[
t_* = 0,\,\,\,\,\,\,
\bar{r}_* > 2\,m,\,\,\,\,\,\,
t'_* = \frac{1}{\sqrt{F(\bar{r}_*)}},\,\,\,\,\,\,
\bar{r}'_* = 0,\,\,\,\,\,\,
\hat{b}_{t*} = 0,\,\,\,\,\,\,
\hat{b}_{\bar{r}*} = - \beta(\bar{r}_*)\,\frac{1}{\sqrt{F(\bar{r}_*)}}.
\]
The $\tilde{g}$-normalization gives  
\begin{equation}
\label{normalizedcg}
F\,(t')^2 - \frac{1}{F}\,(\bar{r}')^2 = 1.
\end{equation}
Solving for $t' > 0$ and inserting the result into
equation (\ref{2xcgschw}) gives 
\begin{equation}
\label{3xcgschw}
0 = \bar{r}'' + \frac{1}{2}\,F,_{\bar{r}} 
- \beta\,\sqrt{F + (\bar{r}')^2}.
\end{equation}
Dividing by the square root and multiplying with $\bar{r}'$ leads to
$d(\sqrt{F + (\bar{r}')^2} - \beta\,\bar{r})/d\,\bar{\tau} = 0$,
whence
\begin{equation}
\label{cgfirstint}
\sqrt{F + (\bar{r}')^2} - \beta\,\bar{r} = \gamma,
\end{equation}
with the constant of integration given by
$\gamma = \sqrt{F(\bar{r}_*)} - \beta(\bar{r}_*)\,\bar{r}_*$. 
It follows that
\begin{equation}
\label{rbardotsign}
\bar{r}' = \pm\,\sqrt{(\gamma + \beta\,\bar{r})^2 - F(\bar{r})},
\end{equation} 
with a sign which depends on the value of $\bar{r}_*$.

Given $\bar{r}(\bar{\tau})$, we obtain $t(\bar{\tau})$ by integrating
$(t')^2 = (\beta\,\bar{r} + \gamma)^2\,F^{-2}$ for the given initial
conditions. However, the integration of $t$ is in general not
particularly interesting, because $t$ does neither extend smoothly
through ${\cal J}^+$ nor through the future horizon.

To study the extension of the conformal geodesics through ${\cal  J}^+$
it is convenient to use the first of the line elements
(\ref{retadvnullSchw}). The conformal geodesic equations then read 
\[
w'' - \frac{1}{2}\,F,_{\bar{r}}\,(w')^2 = -
\beta\,w',\,\,\,\,\,\,
\bar{r}'' + \frac{1}{2}\,F\,F,_{\bar{r}}\,(w')^2 +
F,_{\bar{r}}\,\bar{r}'\,w' = \beta\,(\bar{r}' +
F\,w').
\]
These equations imply for $\bar{r}$ the equations obtained above. The
normalization of the tangent vector reads
$F\,(w')^2 + 2\,w'\,\bar{r}' = 1$. Solving for $w'$ and requiring
$w' > 0$ on $\tilde{S}$ gives
\begin{equation}
\label{wequation}
w'
= \frac{1}{F}\,(\sqrt{F + (\bar{r}')^2} - \bar{r}')
= \frac{1}{\sqrt{F + (\bar{r}')^2} + \bar{r}'},
\end{equation}
which has to be integrated with the initial conditions
$w_* = - \bar{r}_* - 2\,m\, \log (\bar{r}_* - 2\,m)$.
Thus $w(\bar{\tau})$ can be obtained by a simple integration once
$\bar{r}(\bar{\tau})$ has been determined.

To study the extension through the horizon, it is convenient to use the
second of the line elements (\ref{retadvnullSchw}). The conformal geodesic
equations then read
\[
v'' + \frac{1}{2}\,F,_{\bar{r}}\,(v')^2 =
\beta\,v',\,\,\,\,\,\,
\bar{r}'' + \frac{1}{2}\,F\,F,_{\bar{r}}\,(v')^2 -
F,_{\bar{r}}\,\bar{r}'\,v' 
= - \beta\,(\bar{r}' - F\,v').
\]
The normalization of the tangent vector gives now
$F\,(v')^2 - 2\,v'\,\bar{r}' = 1$, which leads to the
equation
\begin{equation}
\label{vequation}
v' = 
\frac{1}{F}\,(\sqrt{F + (\bar{r}')^2} + \bar{r}'),
\end{equation} 
which has to be integrated with the initial conditions
$v_* = \bar{r}_* + 2\,m\, \log (\bar{r}_* - 2\,m)$.

\subsubsection{Conformal geodesics on which $\bar{r}$ is constant}

The explicit solution of the key equation (\ref{rbardotsign}) requires
the discussion of three different cases. We first consider the
borderline solution.

A change of the sign in (\ref{rbardotsign}) should occur near conformal
geodesics along which $\bar{r}$ is constant. Requiring
$\bar{r}' = 0$ in (\ref{3xcgschw}) gives 
$1/2\,F' = \beta\,\sqrt{F}$.
With $\beta$ given by (\ref{2beta})
this condition has the solution
\[
\bar{r}_* = \hat{r} \equiv \frac{5}{2}\,m 
\quad\mbox{resp.}\quad
r = r_{\pm} \equiv \frac{3 \pm \sqrt{5}}{4}\,m.
\]
Inserting $\bar{r}' = 0$ into (\ref{normalizedcg}), using 
(\ref{taubaroftau}) with $\tau_* = 0$ , $\bar{\tau}_* = 0$
, and observing (\ref{2Theta}), (\ref{2beta}) with
$r = r_+$ gives on the conformal geodesics on
which $\bar{r} = \hat{r}$
\[
t = \frac{\bar{\tau}}{\sqrt{F(\hat{r})}} =
 \frac{25}{4}\,m\,\log \left(\frac{2\,\Theta_* + \beta\,\tau}
{2\,\Theta_* - \beta\,\tau}\right),
\]
and thus $t \rightarrow \infty$ as
$\tau \rightarrow \tau_{i} \equiv \frac{2\,\Theta_*(r_+)}{\beta(r_+)}$.

It follows now from equations (\ref{3xcgschw}), (\ref{rbardotsign})
that each of the conformal geodesics specified by our data
falls into one of the following four classes: (i) the conformal geodesics
passing through points with 
$r = \frac{m}{2}$, which coincide with metric geodesics in the
Schwarzschild-Kruskal space-time and approach the singularity, (ii)
the conformal geodesics passing through points with 
$r = r_{\pm}$, which are tangent to the static Killing field
$\partial_t$ and approach time-like infinity for the finite value
$\tau_{i}$ of their parameters,
(iii)  the conformal
geodesics passing through points with 
$0 < r < r_{-}$ or with 
$r_{+} < r < \infty$, for which $\bar{r}$
is monotonically increasing,
(iv) the conformal geodesics passing through points with 
$r_{-} < r < \frac{m}{2}$ or with
$\frac{m}{2} < r < r_{+}$, for which $\bar{r}$
is monotonically decreasing.
Though it will be seen below that the integration procedure for
(\ref{rbardotsign}) depends on the classes specified
above, it is clear that our data, which are smooth on
$\tilde{S}$, determine a smooth congruence of conformal geodesics
which is free of caustics near $\tilde{S}$.

\subsubsection{Conformal geodesics on which $\bar{r}$ is increasing}

Because of the symmetry of the data it is sufficient to discuss the case
$r_+ < r < \infty$. Under this assumption the ``$+$'' sign holds in
(\ref{rbardotsign}). The radicand in this equation factorizes as
\begin{equation}
\label{intfact}
(\gamma + \beta\,\bar{r})^2 - F(\bar{r})
= \frac{\beta^2}{\bar{r}}\,(\bar{r} - \bar{r}_*)\,
(\bar{r} - \alpha_+)\,(\bar{r} - \alpha_-).
\end{equation}
Because $\gamma = - \sqrt{F(\bar{r}_*)}$, it follows that 
\[
\alpha_{\pm} = \pm \,\alpha
\quad\mbox{with}\quad
\alpha = \sqrt{\frac{m\,\bar{r}_*}{2\,F(\bar{r}_*)}}.
\]
Since $\alpha < \bar{r}_*$ and $\alpha \rightarrow \hat{r}$ as
$\bar{r}_* \rightarrow \hat{r}$, the polynomial on the right hand side of
(\ref{intfact}) has under our assumption three different zeros while in
the limit above two zeros will coincide and the integration procedure
needs to be changed. 

The use of the notation
\[
t \equiv \left(\frac{(\alpha + \bar{r})\,\bar{r}_*}
{(\alpha + \bar{r}_*)\,\bar{r}}\right)^{\frac{1}{2}},
\,\,\,\,\,\,\,\,
k \equiv  \left(
\frac{1}{2}(1 + \frac{\alpha}{\bar{r}_*})\right)^{\frac{1}{2}}
= \left(\frac{1}{2}
(1 + \sqrt{\frac{m}{2\,\bar{r}_* - 4\,m}})\right)^{\frac{1}{2}}
,\,\,\,\,\,\,\,\,
e \equiv \sqrt{2}\,k,
\]
in (\ref{rbardotsign}) gives, after an integration, 
\[
\beta\,\bar{\tau} =
\frac{2}{e}\,\sqrt{(e^2 - 1)\,(e^2 - k^2)}\,
\int\limits_{\,1}^{\,\frac{1}{e}\sqrt{1 + \frac{\alpha}{\bar{r}}}}
\frac{d\,t}{
(1 - e^2\,t^2)\,
\sqrt{(1 - t^2)\,(1 - k^2\,t^2)}},
\]
where, by our assumptions, $1 < e < \sqrt{2}$, 
$\frac{1}{\sqrt{2}} < k < 1$.
It is well known how to express the elliptic integral of the third kind
which occurs above in terms of theta functions
(for all manipulations involving elliptic integrals, elliptic functions,
theta functions, etc. we refer to \cite{lawden}). The substition
$t = sn(u,k)$ with Jacobi's elliptic function $sn$ yields
\begin{equation}
\label{explbaseint}
\beta\,\bar{\tau}
= 
\frac{2}{e}\,\sqrt{(e^2 - 1)\,(e^2 - k^2)}\,
\int\limits_{\,K}^{u}
\frac{d\,v}{(1 - e^2\,sn^2 v)}
\end{equation}
\[
= - 2\,Z(c)\,(u - K) 
+ \log\,\left(\frac{\theta_1(\frac{\pi}{2\,K}(u + c))}
{\theta_1(\frac{\pi}{2\,K}(u - c))}\right).
\]
Here $Z$ denotes Jacobi's zeta function
\[
Z(u) = \frac{d}{d\,u}\,(\log\,\theta_4(\frac{\pi}{2}\,\frac{u}{K}))
= \frac{2\,\pi}{K}\,\sin (\pi\,\frac{u}{K})
\,\sum_{n = 1}^{\infty}
\frac{q^{2n - 1}}{1 - 2\,q^{2n - 1}\,\cos (\pi\,\frac{u}{K})
+ q^{4n - 2}}.
\]
Thus $Z$ is an analytic function on the real line with
$Z(u) > 0$ for $0 < u < K$ and $Z(0) = Z(K) = 0$.
The theta functions are given in their product representations by 
\[
\theta_1(z) = 2\,q^{\frac{1}{4}}\,\sin\,z\,\prod_{n = 1}^{\infty} 
(1 - q^{2\,n})\,(1 - 2\,q^{2\,n}\,\cos\,2\,z + q^{4\,n}),
\]
\[
\theta_2(z) = 2\,q^{\frac{1}{4}}\,\cos\,z\,\prod_{n = 1}^{\infty} 
(1 - q^{2\,n})\,(1 + 2\,q^{2\,n}\,\cos\,2\,z + q^{4\,n}),
\]
\[
\theta_4(z) = \prod_{n = 1}^{\infty} 
(1 - q^{2\,n})\,(1 - 2\,q^{2\,n - 1}\,\cos\,2\,z + q^{4\,n - 2}),
\]
with $q = \exp(-\pi\,\frac{K'}{K})$. The constant $K > 0$ is the
complete integral of the first kind
\[
K(k) = \int\limits_0^1
\frac{d\,t}{\sqrt{(1 - t^2)\,(1 - k^2\,t^2)}}, 
\]
$K' = K(k')$ with $k^{'2} = 1 - k^2$, and $c$ is the (unique) number
in the open interval $]0, K[$ satisfying $sn\,(c, k) = 1/e$.

With $\tau_* = 0$, $\bar{\tau}_* = 0$, (\ref{2Theta}), and (\ref{2beta})
we get from equation (\ref{taubaroftau}) 
\begin{equation}
\label{goftau}
\beta\,\bar{\tau} = - \log \,g(\tau)
\quad\mbox{with}\quad
g(\tau) \equiv \frac{2\,\Theta_* - \beta\,\tau}
{2\,\Theta_* + \beta\,\tau}.
\end{equation}
Setting now $x = K - u$ to exhibit the symmetries of the
solution, we obtain 
\[
\tau = G(x, k) \equiv
\frac{2\,\Theta_*}{\beta}\,\,
\frac{\theta_2(\frac{\pi\,(c - x)}{2\,K})\,e^{x\,Z(c)} -
\theta_2(\frac{\pi\,(c + x)}{2\,K})\,e^{- x\,Z(c)}}
{\theta_2(\frac{\pi\,(c - x)}{2\,K})\,e^{x\,Z(c)}\, +
\theta_2(\frac{\pi\,(c + x)}{2\,K})\,e^{- x\,Z(c)}}.
\]
Since the denominator of the function $G$ is positive in an open 
neighbourhood of the closed interval $[ - (K - c), K - c]$, the
function $G$ is analytic there. While
$\bar{\tau} \rightarrow \infty$ as $u \rightarrow c$ and 
$\bar{\tau} \rightarrow - \infty$ as $u \rightarrow 2\,K - c$
by (\ref{explbaseint}), we have
now $\tau \rightarrow \pm \frac{2\,\Theta_*}{\beta}$, whence 
$\bar{\tau} \rightarrow \pm \infty$, as $x \rightarrow \pm (K - c)$.
It follows from the ODE satisfied by $\bar{r}$ that we can solve
the equation above for $x(\tau)$ with 
$\tau \in \,] - \frac{2\,\Theta_*}{\beta}, \frac{2\,\Theta_*}{\beta}[$. 
A direct calculation shows that $G'(x, k) > 0$ at $x = \pm (K - c)$.
This implies that $x(\tau)$ extends as an analytic function into an open
neighbourhood of
$[ - \frac{2\,\Theta_*}{\beta}, \frac{2\,\Theta_*}{\beta}]$

Inserting now $x(\tau)$ into the equation $t = sn(K - x,k)$, we
finally get
\begin{equation}
\label{rlargefinform}
\frac{\bar{r}_*}{\bar{r}} = 1
- \frac{2\,(1 - k^2)\,k^2}{2\,k^2 - 1}\,
\frac{sn^2 (x(\tau))}{1 - k^2\,sn^2 (x(\tau))}.
\end{equation}

In the limit when $r \rightarrow r_+$ we have $k \rightarrow 1$ and thus
$sn (u, k) \rightarrow \tanh u$. The formula (\ref{rlargefinform})
thus suggests that the solution $\bar{r}(\tau, r)$ of
(\ref{rlargefinform}) approaches the constant solution $\bar{r} = \hat{r}$
in that limit, as we know already by general arguments. However, since $K
\rightarrow \infty$
while $c \rightarrow 1/2\,\log((\sqrt{2} + 1)/(\sqrt{2} - 1))$
as $k \rightarrow 1$, the precise behaviour of the
solution in that limit requires a quite careful
discussion involving also the behaviour of $G(x, k)$. 

The right hand side
of  (\ref{rlargefinform}) is an analytic function
of $\tau$ in an open intervall containing 
$[- \tau_{scri}(r), \tau_{scri}(r)]$, with
$\tau_{scri}(r) \equiv \frac{2\,\Theta_*}{\beta}
= \frac{r}{(r + \frac{m}{2})\,(r - \frac{m}{2})}$.
It vanishes precisely at
the points $\tau = \pm \tau_{scri}$ at which the conformal
factor (\ref{2Thetaoftau}) vanishes. Furthermore, it depends analytically
on $r$ for $r_+ < r < \infty$. It follows that 
\begin{equation}
\label{barrTpos}
\bar{r}(r, \tau)\,\Theta(r, \tau)
\quad\mbox{is positive and analytic for}\quad
r \in ]r_+, \infty[,\,\,
\tau \in [- \tau_{scri}(r), \tau_{scri}(r)].
\end{equation}
Using the coordinate $z = \frac{1}{\bar{r}}$ in the first of the line
elements (\ref{retadvnullSchw}) and rescaling with the conformal factor 
$\Omega = z$ gives the smooth conformal representation
\begin{equation}
\label{Schwarzwzform}
\Omega^2\,\tilde{g} = z^2\,(1 - 2\,m\,z)\,d\,w^2
- 2\,d\,w\,d\,z - d\,\sigma^2,
\end{equation}
of the Schwarzschild metric which extends analytically through future
null infinity, given here by ${\cal J}^+ = \{z = 0\}$.  

Integrating (\ref{wequation}) with $\bar{r}$ given by
(\ref{rlargefinform}) and writing the solution in terms of $z$, we obtain
our conformal geodesics in the form 
\begin{equation}
\label{wzcgform}
\tau \rightarrow (w(\tau, r),\,\,z(\tau, r)).
\end{equation}
From the discussion above it is clear that the first of the functions on
the right hand side is an analytic function of both variables for 
$r \in ]r_+, \infty[$, $\tau \in [- \tau_{scri}(r), \tau_{scri}(r)]$. 
We show that this holds
true also for the second function. It is clear that the function $w$ is
analytic near $\tilde{S}$. Parametrizing $w$ in terms of $z$, we
obtain from (\ref{rbardotsign}), (\ref{wequation}) 
\[
\frac{d\,w}{d\,z} = - 
\left( \beta^2\,(1 - \bar{r}_*\,z)\,(1 - \alpha^2\,z^2)
+ (\beta + \gamma\,z)\,
\sqrt{ \beta^2\,(1 - \bar{r}_*\,z)\,(1 - \alpha^2\,z^2)}\right)^{-1}.
\]
The assertion now follows because the function on the right hand side is
analytic for $z$ in an open interval of the form 
$] - \epsilon, \frac{1}{\bar{r}_*}[$ with some $\epsilon > 0$.

It follows in particular that there exists a smooth function
$\hat{w}(r)$, with $r \in ]r_+, \infty[$, such that 
$w(\tau, r) \rightarrow \hat{w}(r)$ 
as $\tau \rightarrow \tau_{scri}$ (or, for symmetry reasons, as 
$\tau \rightarrow - \tau_{scri}$). We show that 
\[
\hat{w}(r) \rightarrow \infty
\quad\mbox{as}\quad r \rightarrow r_+,
\,\,\,\,\,\,\,\,\,
\hat{w}(r) \rightarrow - \infty
\quad\mbox{as}\quad r \rightarrow \infty.
\]
The first assertion follows from the observation that 
\[
w = - \hat{r} - 2\,m\, \log (\hat{r} - 2\,m)
+ \frac{25}{4}\,m\,\log \left(\frac{2\,\Theta_* + \beta\,\tau}
{2\,\Theta_* - \beta\,\tau}\right),
\]
along the conformal geodesic with $\bar{r} = \hat{r}$ and the fact
that the solutions are jointly smooth in the initial data and the
parameter. The second assertion follows from a comparison of $w$ with
the solution to  
\[
\frac{d\,u}{d\,z} = - 
\left(\beta^2\,(1 - \bar{r}_*\,z)\,(1 - (\frac{\alpha}{\bar{r}_*})^2)
+ \frac{1}{2}\,\sqrt{(1 - \bar{r}_*\,z)\,(1 -
(\frac{\alpha}{\bar{r}_*})^2)}\right)^{-1},
\]
which satisfies $u = w$ at $z = \frac{1}{\bar{r}_*}$. Since 
$0 \ge \frac{d\,w}{d\,z} \ge \frac{d\,u}{d\,z}$ 
for $z \in [0, \frac{1}{\bar{r}_*}]$, 
it follows that 
the value $\hat{u}(r)$ of $u$ at $z = 0$ gives an upper estimate for 
$\hat{w}(r)$. Since the direct integration gives $\hat{u}(r)$ with
$\hat{u}(r) \rightarrow - \infty$ as $r \rightarrow \infty$, the
assertion follows.

\subsubsection{Conformal geodesics on which $\bar{r}$ is decreasing}

It is sufficient to discuss the case $\frac{m}{2} < r < r_{-}$.
Now (\ref{rbardotsign}) must hold with the ``$-$'' sign. The
function $\alpha$ in the factorization (\ref{intfact}) then* satisfies
$\bar{r}_* < \alpha$ and $\alpha \rightarrow \infty$ as
$\bar{r}_* \rightarrow \frac{m}{2}$ while
$\alpha \rightarrow \hat{r}$ as $\bar{r}_* \rightarrow \hat{r}$.
If we set
\[
t \equiv \left(\frac{1}{2}\,(1 + \frac{\alpha}{\bar{r}})
\right)^{\frac{1}{2}},\,\,\,\,\,\,\,\,
k \equiv 
\left(\frac{1}{2}\,(1 + \frac{\alpha}{\bar{r}_*})\right)^{- \frac{1}{2}},
\,\,\,\,\,\,\,\,
e \equiv \sqrt{2},
\]
such that $e > 1$, $0 < k <1$, the integration of
(\ref{rbardotsign}) yields 
\[
\beta\,\bar{\tau} = \sqrt{2\,(2 - k^2)}\,
\int \limits_{\frac{1}{k}}^{\frac{1}{e}\,
\sqrt{1 + \frac{\alpha}{\bar{r}}}}
\frac{d\,t}
{(t^2\,e^2 - 1)\,\sqrt{(t^2\,k^2 - 1)\,(t^2 - 1)}}.
\]
The substitution $k\,sn(u,k) = 1/t$ and 
$c \in ]0, K[$ such that $sn(c,k) = 1/e$ give
\[
\beta\,\bar{\tau} 
= k^2\,\sqrt{\frac{2 - k^2}{2}}\, 
\int \limits_u^{K}
\frac{sn^2v}{1 - \frac{k^2}{2}\,sn^2v}\,d\,v
= - 2\,Z(c)\,(u - K) +
\log 
\frac{\theta_4\left(\frac{\pi\,(u + c)}{2\,K}\right)}
{\theta_4\left(\frac{\pi\,(u - c)}{2\,K}\right)}.
\]
Setting $x = K - u$ and observing (\ref{goftau}), we finally obtain
\begin{equation}
\label{tauxrel}
\tau = F(x, k) \equiv \frac{2\,\Theta_*}{\beta}\,
\tanh \left(k^2\,\sqrt{\frac{2 - k^2}{2}}\,
\int\limits_0^x \frac{1 - sn^2 v}{2 - k^2\,(1 + sn^2v)}\,d\,v
\right).
\end{equation}
The function $F(x)$ is real analytic on $\mathbb{R}$ with $F'(x) \ge 0$
and $F'(x) = 0$ iff $x = (2\,m + 1)\,K$ with $m \in \mathbb{Z}$. Thus
(\ref{tauxrel}) can be solved to obtain an analytic function $x(\tau)$
which maps the interval $]- \tau_s, \tau_s[$, with  
$2\,\Theta_*/\beta < \tau_s \equiv G(K,k)$, 
diffeomorphically onto $]- K, K[$. The solution can then be written
\begin{equation}
\label{rsmallfinform}
\bar{r} = \bar{r}_*\, 
\frac{(2 - k^2)\,(1 - sn^2 (x(\tau)))}{2 - k^2\,(1 + sn^2(x(\tau)))}.
\end{equation}

From the discusssion above it follows that 
$\bar{r}(\tau) \rightarrow 0$ and 
$d\,\bar{r}/d\,\tau \rightarrow \mp \infty$ as $\tau \rightarrow
\pm \tau_s$. Thus in this case there does not exists a smooth
(though a continuous) extension of $\bar{r}(\tau)$ beyond its physical
domain. The limit $r \rightarrow r_+$ implies $k \rightarrow 1$.
In particular, it follows from (\ref{tauxrel}), (\ref{rsmallfinform}) that
the right hand side of (\ref{rsmallfinform}) goes in this limit to
$\hat{r}$ for constant $\tau$ and $\tau_s \rightarrow \tau_i$. 

The limit $r \rightarrow \frac{m}{2}$
implies $k \rightarrow 0$. Since $\beta = 0$ at $r = \frac{m}{2}$, the
expression for $\tau_s$ appears to give a nonsensical result at that
point. However, we have 
$2\,\Theta_*\,k^2/\beta \rightarrow 2/m$. Expanding the right hand side
of (\ref{tauxrel}) and observing that $K(0) = \pi/2$ thus 
gives $\tau_s \rightarrow \pi/4\,m$ as $r \rightarrow m/2$,
consistent with the fact the conformal geodesics approach in the limit
metric geodesics with length
$\bar{\tau} = \frac{\tau_s}{\Theta_*(\bar{r}_* = 2þm)} = m\,\pi$. 

To follow the conformal geodesics through the horizon, one has to
integrate for given $\bar{r}(\tau)$ the equation
\begin{equation}
v' = 
\frac{\gamma + \beta\,\bar{r}
- \sqrt{(\gamma + \beta\,\bar{r})^2 - F(\bar{r})}}{F(\bar{r})},
\end{equation}
whose right hand side defines a positive analytic function of
$\bar{r}$ for $\bar{r} > 0$. The conformal geodesics are then obtained 
in the form 
\begin{equation}
\label{vbarrcgform}
\tau \rightarrow x(\tau, r) = (v(\tau, r),\,\,\bar{r}(\tau, r)),
\end{equation}
where the function on the right hand side are analytic for 
$(\tau, r)$ with $r \in ]\frac{m}{2}, r_+[$, $\tau \in [0, \tau_s(r)[$.

\subsection{Analytic coordinates covering the Schwarzschild-\\
Kruskal space-time and its null infinity}

If we set now $y^1 = r$, $y^2 = \phi$, 
$y^3 = \theta$ on $\tilde{S}$, drag these coordinates along with
the congruence of conformal geodesics constructed in the previous
section, and set $y^0 = \tau$, we obtain a smooth coordinate system
near $\tilde{S}$. The purpose of the following discussion is to
establish the following result (we ignore here the coordinate
singularity arising from the use of coordinates on $S^2$ which can
easily be removed):\\
 
{\it The conformal Gauss system $y^{\nu}$ defines a smooth global
coordinate system on the Schwarz\-schild-Kruskal space-time which
extends smoothly to null infinity.}\\

It remains to be
shown that the conformal geodesics of the congruence do not
develop caustics. Because of the symmetry of the functions
(\ref{SchwKrusk}) and of our family of curves under $s \rightarrow - s$,
$\rho \rightarrow - \rho$, it is sufficient, to control the behaviour
of the congruence on the subset
$\{s \ge 0, \rho \ge 0\}$ of the Schwarzschild-Kruskal space-time.
Since the coordinates $y^{\mu}$ are regular near $\tilde{S}$, they
are known to be regular in particular near the set 
$\{s = 0, \rho = 0\}$. Therefore we can work with the coordinates 
$w$, $\bar{r}$ or $z$, $\bar{r}$ or $v$, $\bar{r}$ respectively. 

Using the functions on the right hand sides of (\ref{wzcgform}) and
(\ref{vbarrcgform}), we can in principle express the corresponding
line elements in terms of the coordinates $y^{\mu}$. Though the
resulting metric coefficients will be smooth where (\ref{wzcgform}) and
(\ref{vbarrcgform}) are analytic, the Jacobian of the transformation
may drop rank at various places and we may end up with a degenerate
representation of the metric. Having available the explicit solution,
we could try to check this by a direct calculation. Since the
explicit calculation will be tedious and, in particular, because this
method will not apply to more general cases, we prefer to employ an
argument which is similar to the analysis of the behaviour of metric
geodesics congruences in terms of the Jacobi equation. 

Consider the transformation provided by (\ref{vbarrcgform}).
We need to show that the tangent vector field 
$\dot{x} = \Theta^{-1}\,\bar{x}' = \Theta^{-1}\,X$  and the connecting
vector field 
$x_{,r} = \bar{x}_{,r} - \dot{x}\,\tau_{,r} = Z - \dot{x}\,\tau_{,r}$ 
are linearly independent on their domain of
definition. In terms of the 2-form (\ref{2heps}) (with $h = F\,d\,v^2 -
2\,d\,v\,d\,\bar{r}$, $x^0 = v$, $x^1 = \bar{r}$) this can be
expressed as the requirement that the invariant
$\Theta^{-1}\,\epsilon_h(X, Z)$ does not vanish. In the case of
(\ref{wzcgform}) the nondegeneracy up to and in fact beyond null
infinity would follow if it could be shown that 
\[
- (\bar{r}\,\Theta)^2\,(\dot{w}\,z_{,r} - \dot{z}\,w_{,r}) \neq 0
\quad\mbox{for}\quad
r_* \in ]r_+, \infty[,\,\,
\tau \in [0, \tau_{scri}(r)].
\]
If $z$ is replaced by $\bar{r}$ again, this
condition translates in terms of  
$x(\tau, r) = (v(\tau, r),\,\,\bar{r}(\tau, r))$ and
(\ref{2heps}) (with 
$h = F\,d\,w^2 + 2\,d\,w\,d\,\bar{r}$, $x^0 = w$, $x^1 = \bar{r}$)  
into the requirement that
\[
0 \neq \epsilon_{\Theta^2 h}(\dot{x}, \,x_{,r})
= \Theta\,\epsilon_h(X, Z),
\]
in the domain given above. Here the factor $\Theta$ is of course most
significant because it vanishes at $\tau_{scri}$. 

The proof that these requirements are met by our transformations
relies on a differential equation satisfied by $\epsilon_h(X, Z)$. To
make use of (\ref{2vacadaptxconfdeviation}), we observe that the
various representations of the Schwarzschild metric with warping
function $f = \bar{r}$ and second metric $k = - d\sigma^2$ give the
value $c = \frac{1}{2}\,f^2\,D_AD^Af = - m$ for the constant in
(\ref{vacwarpRiem}). It follows 
\[
D_X\,\epsilon_h(X, Z) = \epsilon_h(D_X\,X, Z) + \epsilon_h(X, D_X\,Z)
= \epsilon_h( - \beta\,\epsilon(X,\,.)^{\sharp}, Z) + 
\epsilon_h(X, D_X\,Z)
\]
\[
= - \beta\,h(X, Z) + 
\epsilon_h(X, D_X\,Z),
\]
and similarly
\[
D_X\,D_X\,\epsilon_h(X, Z) =
\]
\[
\beta^2\,\epsilon_h(X, Z) - 2\,\beta\,h(X, D_X\,Z)
+ \epsilon_h(X, D_X\,D_X\,Z)
\]
\[
= (\beta^2 + \frac{2\,m}{\bar{r}^3})\,\epsilon_h(X, Z)
+ D_Z\,\beta,
\]
where in the second equation $D_X\,Z = D_Z\,X$,
$h(X, X) = 1$, and  (\ref{2vacadaptxconfdeviation}) have been used. 
The invariant $\epsilon_h(X, Z)$ thus satisfies the ODE
\[
D_X\,D_X\,\epsilon_h(X, Z) -
(\beta^2 + k)\,\epsilon_h(X, Z)
= D_Z\,\beta,
\]
with $k = \frac{2\,m}{\bar{r}^3}$, and on $\tilde{S}$ the initial
conditions
\[
\epsilon_h(X, Z)_* = \frac{(r + \frac{m}{2})^2}{r^2} > 1,
\,\,\,\,\,\,
(D_X\,\epsilon_h(X, Z))_* = 0 
\quad\mbox{for}\quad r \ge \frac{m}{2}.
\]
The quantity $D_Z\beta$, which is constant along the conformal
geodesics, is given by
\[
D_Z \beta = \beta_{,r} = - 2\,\frac{r^2 -
2\,m\,r +
\frac{m^2}{4}}{(r + \frac{m}{2})^4}.
\]
It vanishes at $\breve{r}_\pm \equiv \frac{2 \pm \sqrt{3}}{2}\,m$, 
where $\bar{r}(\breve{r}_\pm) = 3\,m > \hat{r}$, and satisfies
\[
D_Z \beta > 0
\quad\mbox{for}\quad
\frac{m}{2} \le r < \breve{r}_+,
\,\,\,\,\,\,\,\,\,\,\,\,
D_Z \beta < 0
\quad\mbox{for}\quad
\breve{r}_+ <  r.
\]

A lower estimate for $\epsilon_h(X, Z)$ can be obtained as follows.
Denote by $u$ and $v$ the solutions to the ODE problems
\[
w'' - (\beta^2 + k)\,w = f,\,\,\,\,\,\,w(0) = 1,\,\,\,\,\,\,
w'(0) = 0,
\]
where $f = 0$ in the case of $u$ and $f = - 1$ in the case
of $v$. Then $u$ is strictly increasing with  
$u \ge \cosh(\beta\,\bar{\tau})$ for $\bar{\tau} \ge 0$, and
$\epsilon_h(X, Z)$ can be given in the form
\begin{equation}
\label{inhomsol}
\epsilon_h(X, Z) = u\,\left(\epsilon_h(X, Z)_* 
+ (1 - \frac{v}{u})\,\beta_{,\chi}
\right).
\end{equation}
Since $(u -  v)'' - (\beta^2 + k)\,(u - v) = 1$ and the function
$u - v$ has vanishing initial conditions at $\bar{\tau} = 0$, it
follows that $ u \ge v$ for $\bar{\tau} \ge 0$. Since $v$ changes its
sign, a further estimate is needed. We derive a representation of
$v$ in terms of $u$. Since $u > 0$ there exists a function $f$ with
$v = f\,u$. The ODE's satisfied by $u$ and $v$ imply for $f$
the equation
\[
f'' = - 2\,\frac{u'}{u}\,f' - \frac{1}{u},
\]
which has because of the initial conditions for $u$ and $v$ the
solution
\[
f = 1 - \int_0^{\bar{\tau}} (\frac{1}{u^2}\,
\int_0^{\tau'} u\,d\tau'')\,d\tau'.
\] 
Since $u$ is strictly increasing, it follows
\[ 
0 \le 1 - \frac{v}{u} 
= \int_0^{\bar{\tau}} (\frac{1}{u^2}\,
\int_0^{\tau'} u\,d\tau'')\,d\tau'
\le  \int_0^{\bar{\tau}} \frac{1}{u^2}\,u\,\tau'\,d\tau' 
\le \int_0^{\bar{\tau}} \frac{\tau'}{\cosh (\beta\,\tau')}\,d\tau' 
\]
\[
\le 2\,\int_0^{\bar{\tau}} \tau'\,e^{- \beta\,\tau'}\,d\tau' 
= \frac{2}{\beta^2}\,(1 - 
(\beta\,\bar{\tau} + 1)\,e^{-\beta\,\bar{\tau}}) 
\quad\mbox{for}\quad
\bar{\tau} \ge 0,
\]
which implies 
\begin{equation}
\label{vuqot-1est}
0 \le 1 - \frac{v}{u} \le \frac{2}{\beta^2}.
\end{equation}

A direct calculation gives 
$- 1 < 2\,\beta^{-2}\,D_Z\,\beta < 0$ for $r > \breve{r}_+$.
It follows that
\[
\epsilon_h(X, Z)_* 
+ (1 - \frac{v}{u})\,\beta_{,\chi} \ge \epsilon_h(X, Z)_*
\quad\mbox{for}\quad \frac{m}{2} < r \le \breve{r}_+,
\]
\[
\epsilon_h(X, Z)_* 
+ (1 - \frac{v}{u})\,\beta_{,\chi} \ge \epsilon_h(X, Z)_*
- 1 = \frac{m\,r_* + \frac{m^2}{4}}{r_*^2}
\quad\mbox{for}\quad \breve{r}_+ < r.
\]
Since (\ref{Thetaoftaubar}) gives under our assumptions
\begin{equation}
\label{normparconfac}
\Theta = \frac{\Theta_*}{\cosh^2(\frac{\beta}{2}\,\bar{\tau})},
\end{equation}
(\ref{inhomsol}) implies that for given $r > \frac{m}{2}$
there is a constant $c > 0$ such that 
\[
\Theta\,\epsilon_h(X, Z) 
\ge  
\Theta_*c\,\frac{
\cosh(\beta\,\bar{\tau})}{\cosh^2(\frac{\beta}{2}\,\bar{\tau})}
\ge \Theta_*c.
\]
In the region covered by (\ref{vbarrcgform}) it suffices of course to
get a lower estimate for $\epsilon_h(X, Z)$, because $\Theta$ is
positive where the conformal geodesics approach the singularity.
On the curves with $\bar{r} = \hat{r}$, which seperate the
domains where (\ref{wzcgform}) and (\ref{vbarrcgform}) are valid, 
$c^2 \equiv \beta^2 + k$ with $c = const. > 0$ and thus
\[
\Theta\,\epsilon_h(X, Z) 
= \Theta\,(\epsilon_h(X, Z)_*\,\cosh (c\,\bar{\tau})
+ c^{-2}\,D_Z\beta\,(\cosh (c\,\bar{\tau}) - 1) 
\]
\[
\ge
\Theta_*\,\epsilon_h(X, Z)_* > 0.
\] 
Since $\beta = 0$
but $k \ge k_* = (2þm)^{-2}$, $D_Z\,\beta = m^{-2}$
on the curves starting with $r = \frac{m}{2}$, it follows that 
$\Theta = \Theta_* = (2þm)^{-2}$ and a similar result is obtained.

\subsection{Global numerical evolution for a class of standard Cauchy
data} 

The methods described above offer a possibility to study in detail 
the complete numerical evolution of three-dimensional space-times
without symmetries which are determined by certain asymptotically flat
Cauchy data. In \cite{corvino} has been shown the existence of smooth,
time-symmetric, asymptotically flat solutions of the vacuum constraint
which coincide with certain given time-symmetric initial data on compact
sets and with Schwarzschild data in a neighbourhood of space-like
infinity.  If these data can be constructed numerically, it is easy to
determine numerically hyperboloidal initial data implied by the Cauchy
data. 

Consider the time symmetric initial data set (\ref{SchwKdat}).
The set $\tilde{S}$ can be embedded by the transformation 
$r = \tan \frac{\chi}{2}$ into the 3-dimensional standard unit sphere 
$S = S^3$. With the conformal factor
$\Theta^S_* = 2\,(1 + r^2)^{-1}\,(1 + \frac{m}{2\,r})^{-2}
= 2\,\cos^2\frac{\chi}{2}\,(1 + \frac{m}{2}\cot \frac{\chi}{2})^{-2}$,
the rescaled metric 
$(\Theta^S_*)^2\,\tilde{h} = - (d\,\chi^2 + \sin^2 \chi\,d\,\sigma^2)$
coincides with the restriction of the standard metric on $S^3$ to the
set
$S \setminus \{i_0, i_{\pi}\}$ , where 
$i_0 = \{\chi = 0\}$, $i_{\pi} = \{\chi = \pi\}$. If we set $m = 0$,
we get Minkowski data and conformally compactified Minkowski data
respectively.

For given $\chi_0$ with $\frac{\pi}{2} < \chi_0 < \pi$ let 
$\xi$ be a smooth function on $\mathbb{R}$ such that $\xi(x) = 0$ for
$x \le \frac{\pi}{2}$, $\xi(x) = 1$ for $x \ge \chi_0$ and such that
$\xi' \ge 0$. We define $\psi \in C^{\infty}(]- \infty, \pi[)$ by
setting it equal to 1 for $x \le \frac{\pi}{2}$ and equal to 
$1 + \xi(x)\,( \cot \frac{x}{2} - 1)$ for 
$\frac{\pi}{2} < x < \pi$. Then $\psi' \le 0$ and
$\frac{1}{2} \le M \equiv \sup_{x < \pi}\frac{d\,\psi}{d\,x}(x)
< \infty$.

Suppose we are given time symmetric initial data on $\mathbb{R}^3$ which
agree with Schwarzschild data of mass $m < \frac{1}{M}$ near space-like
infinity and are such that the metric $\tilde{h}$, suitably written in
terms of the coordinates $\phi, \theta, \chi$ on 
$S$, satisfies
\begin{equation}
\label{Schwatinfty}
\tilde{h} = - 
\frac{(1 + \frac{m}{2}\cot \frac{\chi}{2})^4}{4\,\cos^4\frac{\chi}{2}}
\,(d\,\chi^2 + \sin^2 \chi\,d\,\sigma^2)
\quad\mbox{for}\quad \chi > \chi_0.
\end{equation}
The conformal factor 
\begin{equation}
\label{famfactor}
\Theta_* = 
\frac{2\,\cos^2\frac{\chi}{2}}{(1 + \frac{m}{2}\,\psi(\chi))^2},
\end{equation}
is smooth on $S \setminus \{i_{\pi}\}$, coincides for $\chi \ge
\chi_0$ with $\Theta^S_*$, has non-vanishing gradient in $S \setminus
\{i_0, i_{\pi}\}$ and goes to zero at $i_{\pi}$. It defines a conformal
compactification of the data such that $h = \Theta^2_*\,\tilde{h}$
coincides with the standard metric of the $3$-sphere for $\chi >
\chi_0$. We choose initial data $b_*$ for the 1-form field which
annihilate the normals of the initial hypersurface and satisfy
$b_* = \Theta_*^{-1}\,d\,\Theta_*$ in $S \setminus \{i_{\pi}\}$. 

Since the time evolution of the data will be Schwarzschild near
$i_{\pi}$, it can be determined there by the methods described above.
It is clear that we can construct a smooth hyperboloidal hypersurface,
which coincides with $S$ on the the set $\{\chi \le \chi_0\}$ and
extends to the future null infinity of the Schwarzschild part. It
should not be difficult to determine the corresponding initial data
for the conformal field equations, possibly by a numerical
integration (as shown in \cite{friedrich:i0}, this reduces to solving a
system of ODE's). Since there are codes available to evolve such data
numerically  (cf. \cite{frauendiener}, \cite{huebner:iplus} and the
references given there) we could in principle calculate their evolution
in time.

If such data, satisfying (\ref{Schwatinfty}) for a fixed $\chi_0$,
can be constructed for sufficiently small mass $m$ they will be
close to corresponding Minkowskian data and the result of 
\cite{friedrich:first} suggests that they will evolve into solutions 
possessing complete past and future null infinities and regular points
$i^{\pm}$ at future and past time-like infinity. If we use a gauge in
which $b_*$ picks up on $S \setminus \{i_{\pi}\}$ a suitable
component in the direction of the tangent vectors of the conformal
geodesics, it will be possible to construct a gauge of the type
considered above which smoothly covers the complete future of the
initial hypersurface as well as
${\cal J}^+ \cup \{i^+\}$. The work in \cite{huebner:iplus} 
suggests that such solutions can be calculated numerically in their
entirety.

The numerical calculation of such space-time is, of course, not an end
in itself. In fact, the solution will have somewhat curious features.
It follows from \cite{friedrich:kannar} that they will have vanishing
Newman-Penrose constants. If the solution admit regular points $i^+$,
it then follows from \cite{friedrich:schmidt} that the rescaled
conformal Weyl tensor will vanish at those points. This situation,
which is more special than the one considered in \cite{huebner:iplus},
suggests that the method of gluing a Schwarzschild end to a given solution
of the constraints produces data of rather restricted radiation content.
However, such calculations will allow us to perform detailed tests of the
code under completely controlled assumptions and to study the robustness
of the code and of the gauge conditions by choosing data with an
increasing value of the mass which eventually yields the time evolution
of a collapsing gravitational field.

\section{Concluding remarks}

We have described in detail the construction of a global system of
conformal Gauss coordinates on the Schwarzschild-Kruskal solution
which extend smoothly and without degeneracy through null infinity.
Furthermore, we have shown that the conformal factor naturally associated
with this system defines a smooth conformal extension of the
Schwarzschild-Kruskal space-time which gives to null infinity a
finite location in the new coordinates which is determined by our
choice of initial data.

We did not try to work out in detail the behaviour of the fields in
the conformal extension constructed here. An analysis of the fields
near space-like infinity can be found in  \cite{friedrich:i0}. The
behaviour of the fields near time-like infinity, which is of
considerable interest for the numerical calculation of space-times,
has still to be investigated.   

There is a property of the Gauss system which we only indicated
but which may turn out to be quite important. While the regularity of a
conformal Gauss systems is essentially decided by its underlying
conformal geodesics (considered here as point sets), there always
exists a huge class of different time slicings based on one and the
same underlying congruence of conformal geodesics. The consequences
of this freedom still have to be explored.

Of course, there are many more such coordinate systems. It would be
interesting to see whether the initial data for congruences of
conformal geodesics which lead to such coordinates can be
characterized in a general way.


\end{document}